\documentclass[a4paper,fleqn,usenatbib]{mnras}
\usepackage{newtxtext,newtxmath}

\usepackage[T1]{fontenc}
\usepackage{ae,aecompl}
\usepackage{graphicx, subfloat}	
\usepackage{amsmath}	
\usepackage{amssymb}	
\usepackage{float}
\usepackage{subfloat}
\usepackage{caption}
\usepackage{subfig}

\title[Triggered Star Formation]{A Case Study of Triggered Star Formation in Cygnus X}

\author[Deb, Kothes, \& Rosolowsky]{
Soumen Deb,$^{1}$\thanks{E-mail: sdeb@ualberta.ca}
Roland Kothes,$^{2,1}$
Erik Rosolowsky,$^{1}$
\\
$^{1}$Department of Physics, University of Alberta, Edmonton, Alberta, Canada T6G 2E1\\
$^{2}$Dominion Radio Astrophysical Observatory, Herzberg Programs in Astronomy \& Astrophysics, National Research Council Canada, \\~~~P.O. Box 248, Penticton, BC V2A 6J9, Canada \\
}

\date{Accepted XXX. Received YYY; in original form ZZZ}

\pubyear{2018}

\begin{document}
\label{firstpage}
\pagerange{\pageref{firstpage}--\pageref{lastpage}}
\maketitle

\begin{abstract}
Radiative feedback from massive stars can potentially trigger star formation in the surrounding molecular gas.
Inspired by the case of radiatively driven implosion in M16 or Eagle Nebula, we analyze a similar case of star formation observed in the Cygnus X region. We present new JCMT observations of $^{13}$CO(3-2) and C$^{18}$O(3-2) molecular lines of a cometary feature  located at 50 pc north of the Cyg OB2 complex that was previously identified in $^{12}$CO(3-2) mapping.  These data are combined with archival H$\alpha$, infrared, and radio continuum emission data, from which we measure the mass to be 110 M$_\odot$. We identify Cyg OB2 as the ionizing source. We measure the properties of two highly energetic molecular outflows and the photoionized rim. From this analysis, we argue the external gas pressure and gravitational energy dominate the internal pressure.  The force balance along with previous simulation results and a close comparison with the case of Eagle Nebula favours a triggering scenario.  
\end{abstract}



 \begin{center}
    \begin{figure*}
               \includegraphics[width=\textwidth]{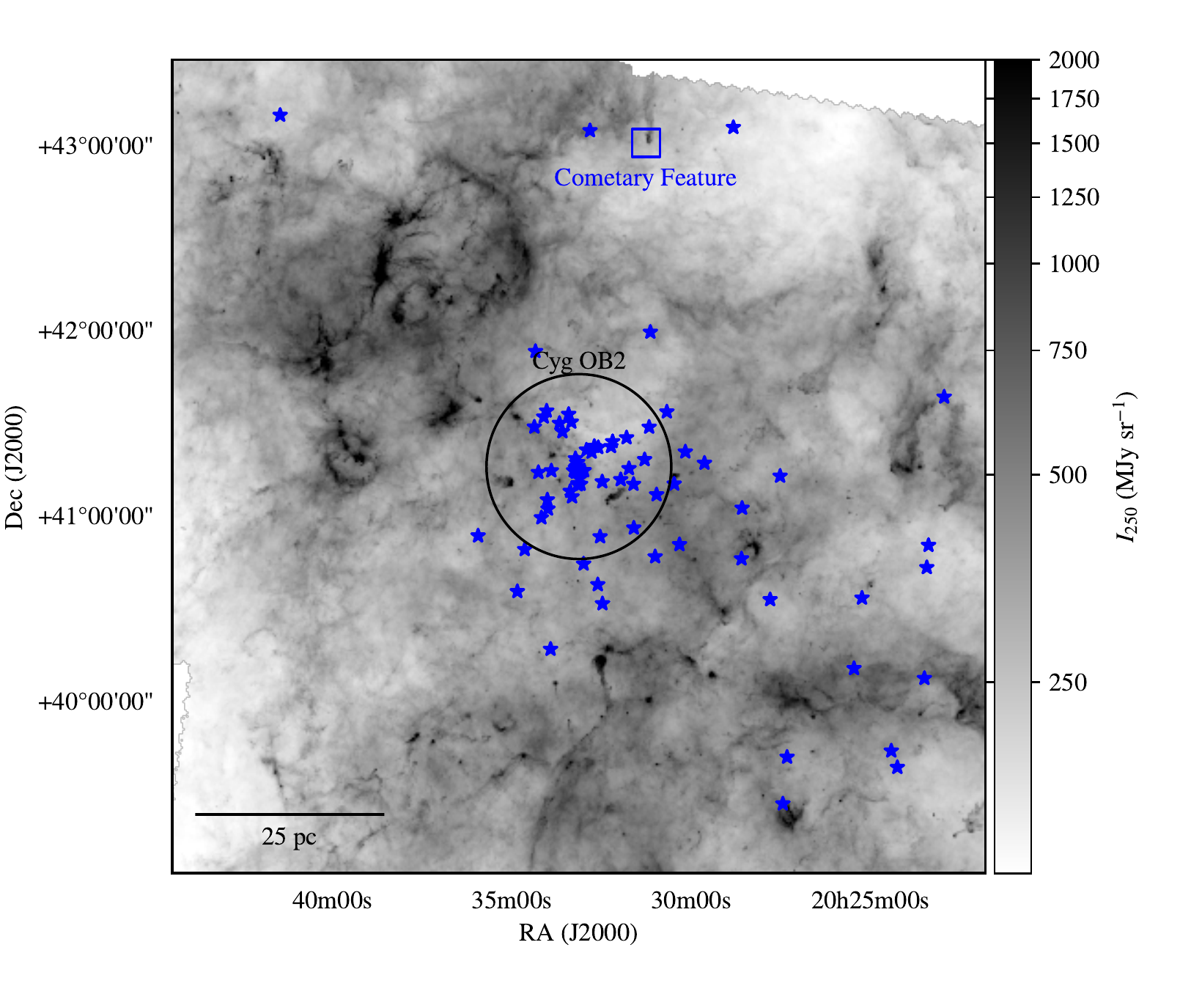}
               \caption{\label{fig:cygx} {\it Herschel} 250 $\mu$m map of the Cygnus OB2 region illustrating the ISM structure in the region. The locations of O stars in the catalog of \citet{B1} are indicated with blue star symbols.  The location of Cyg OB2 is highlighted as is the location of the cometary feature we are studying.  The scale bar on the figure assumes a distance of 1.4 kpc.}
    \end{figure*} 
 \end{center}
\section{Introduction}
Cygnus X is among the most active star forming regions in our Galaxy. It contains hundreds of distinct {\sc H~ii} regions, numerous Wolf-Rayet and O-type stars, and several OB associations with Cygnus OB2 at the heart of the region (Figure~\ref{fig:cygx}). Cygnus X was long considered to be the local spiral arm, the Orion Spur, seen end-on with superimposed emission from objects along a large range of distances \citep[e.g.,][]{wendker91}. This is partly correct, but as shown by \citet{gott12}, there are three major
concentrations of material towards the centre of Cygnus~X. The great Cygnus Rift, responsible for the strong absorption in this direction is located at a distance of 500 to 800~pc and is the only major component along this line of sight that is following normal Galactic rotation \citep{gott12}. The second component, which is connected to star forming region W75N, is located at a distance of 1 to 1.8~kpc. The distance to W75N was later refined by \citet{rygl12} to 1.3~kpc with trigonometric parallax measurements. This component contains material at a high positive local-standard-of-rest (LSR) velocity, which is not allowed assuming a flat rotation curve for our galaxy. This implies that all the material related to W75N was pushed away from us by some force, indicating a triggered origin for this star formation complex. The third component found by \citet{gott12} is related to DR21 at a distance of 1.5 to 2.5~kpc. This was later refined by \citet{rygl12} to 1.5~kpc. Material connected to this component shows an unusually low LSR velocity, in the low negatives, which in this direction of our galaxy would indicate a distance beyond 2.5~kpc. This layer must have been pushed towards us by some force, again, indicating a triggered origin. 

In a JCMT CO(3-2) pilot study of Cygnus~X, \citet{gott12} detected 47 molecular outflows; 27 of these were previously unknown and suggested widespread sequentially-triggered star formation. Most of those outflows and the possibly triggered star formation are related to the aforementioned layers found at unusual LSR velocities. One feature, discovered in the pilot study by \citet{gott12}, is in the shape of a comet with two molecular outflows inside, which is suggestive of triggered star formation for the protostars. This object is found north of Cygnus OB2 (Figure \ref{fig:cygx}). There is a chain of 6 protostellar sources inside this cometary structure \citep{protostars}, two of which produce the two distinctive molecular outflows. The outflows are identified as G81.424+2.140 and G81.435+2.147.  These outflows are also visible in shocked H$_2$ emission in the infrared observations of \citet{makin18}, who also cite the region as a candidate for investigation in the context of triggered star formation. The cometary feature shows a bright limb of emission around the ``head'' of the comet in both H$\alpha$ and the radiocontinuum, implying strong photodissociation and ionization of the molecular gas. Here we use the wide-area CO observations of \citet{gott12}, new data from the James Clerk Maxwell Telescope, and a wealth of archival information to explore this feature as an exemplar of star formation in the neighbourhood of high mass stars.  This work focuses on exploring the interplay between the strong radiation field and winds of the massive stars and how these affect the local star formation in this region.

The question of whether the star formation in this feature has been triggered by external effects is natural given the location of the feature on the edge of the ionized region that is centred on the Cygnus OB2 association of young, high mass stars. The pillar-like geometry of the feature is frequently cited as being indicative of triggered star formation \citep{dale15}, drawing analogy to the well studied ``Pillars of Creation'' in M16 \citep{hester96}. Such features are thought to be formed where a dense region of a molecular cloud casts a shadow with respect to an ionizing radiation field.  The ionizing radiation field can drive an implosion of the dense region at the tip of the forming cometary feature \citep{lefloch94, klein94} and the imploding cloud can then go on to form stars.  Hence, the mere presence of cometary features are often taken as evidence of triggering \citep[][and extensive references therein]{dale15}. 

It is difficult to explicitly identify a region as having had its star formation triggered, mostly because observations lack the ability to evaluate how star formation would have proceeded in a given region under the alternative hypothesis, namely the lack of the strong local ionizing radiation field. Simulations, however, enjoy the ability to actively conduct experiments, studying star formation with and without feedback.  Observations can only make a case for some physical effect locally increasing the star formation rate.  However, the expected outcomes of star formation are only well-defined on large scales where the star formation law and initial mass function become well defined \citep{kennicutt12}. Thus, the case for triggering of star formation on small scales ($<10$~pc) are particularly fraught \citep{dale15} and must be approached with a careful study.

In this work, we examine the cometary feature in detail using new observations of CO isotopologues to characterize the outflows in this region.  We present these new data in Section \ref{sec:obs} and calculate the physical properties of the feature as as whole as well as the outflows in Section \ref{sec:results}.  In Section \ref{sec:discuss}, we discuss the implications of these measurements in the context of triggering.


\section{Observations}
\label{sec:obs}
We observed $^{13}$CO (3-2) and C$^{18}$O (3-2) emission with the James Clerk Maxwell Telescope (JCMT), located at the summit of Mauna Kea in Hawai'i, using the Heterodyne Array Receiver Program (HARP) instrument and the Auto Correlation Spectral Imaging System (ACSIS) spectrometer (project code M11AC10). HARP consists of 16 independent receptors in the focal plane of the instrument, separated by $30''$ on the sky.  We observed the region using the JIGGLE4 scan pattern, which samples a grid of positions covering the region filling in the positions between the individual receptors at sub-Nyquist sampling.  We collected sixteen jiggle maps at two offset pointing centres to make a deep map of the outflows.  The receiver system observed two bands centred at 330.58 GHz and 329.33 GHz for $^{13}$CO(3-2) and C$^{18}$O (3-2) transitions respectively with a velocity resolution of 0.055 km~s$^{-1}$. Over the course of the observations (July 2011), the mean atmospheric opacity at 225 GHz was $\tau_{225} = 0.041$, corresponding to Band 2 weather in JCMT operations. We reduced the data using the {\sc Starlink} software package \citep{STARL} and the standard JCMT reduction recipes for narrow emission line data.  We then combined the data into a mosaic (position-position-velocity) data cube for each molecular species with central coordinates at RA(J2000) = $20^{\rm h}31^{\rm m}12.5^{\rm s}$, Dec(J2000) = $+43^{\circ}05^{'}33.8^{''}$. We used nearest-neighbour gridding to map the positions of the spectra into the final data cube. We fit and subtracted a third-order baseline from each position in the data cubes, using emission-free regions defined by eye in the $^{13}$CO data. The beam size for the observations is $15.2''$. Since the data are sampled on a $7.5''$ rectilinear grid, only spatial scales larger than $2\sqrt(2)\cdot 7.5''=21.2''$ are fully sampled in all directions. The small-scale spatial distribution of the emission is systematically uncertain, but the primary conclusions of this study rest primarily on the integrated spatial properties of the molecular emission, which will be less affected. We corrected the emission to the $T_\mathrm{MB}$ scale using the observatory values for the beam efficiency: $\eta_{\mathrm{MB}}=0.64$. The noise level in the final cubes is 130 mK in a 0.055 km~s$^{-1}$ channel, when measured on the $T_{\mathrm{MB}}$ scale.  In this work, we also use the $^{12}$CO(3-2) data from \cite{gott12}, which we align and re-sample to be on the same coordinate grid as the $^{13}$CO and C$^{18}$O data.

\begin{figure*}
    \centering
\includegraphics[width=0.49\textwidth]{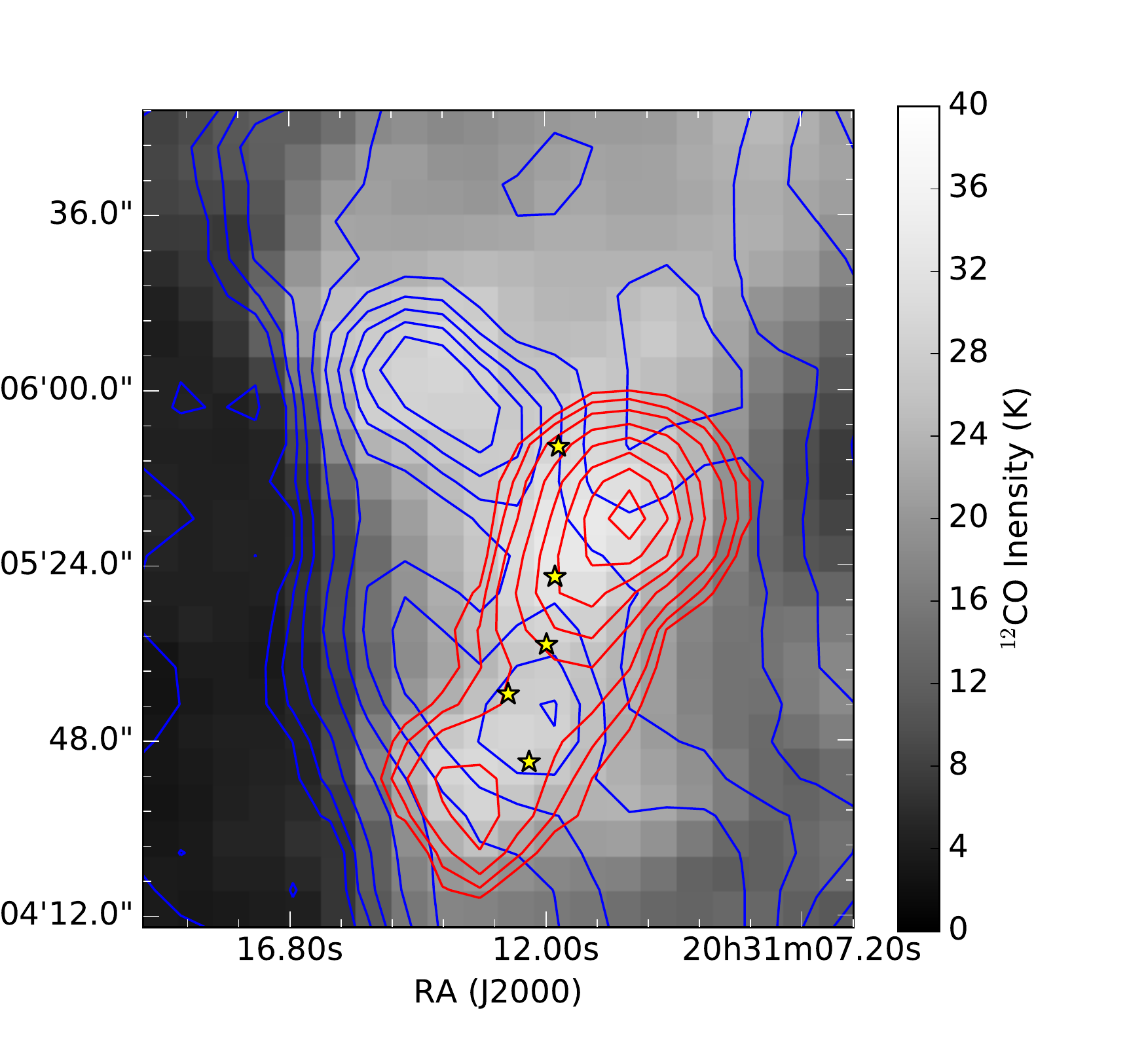}
\includegraphics[width=0.49\textwidth]{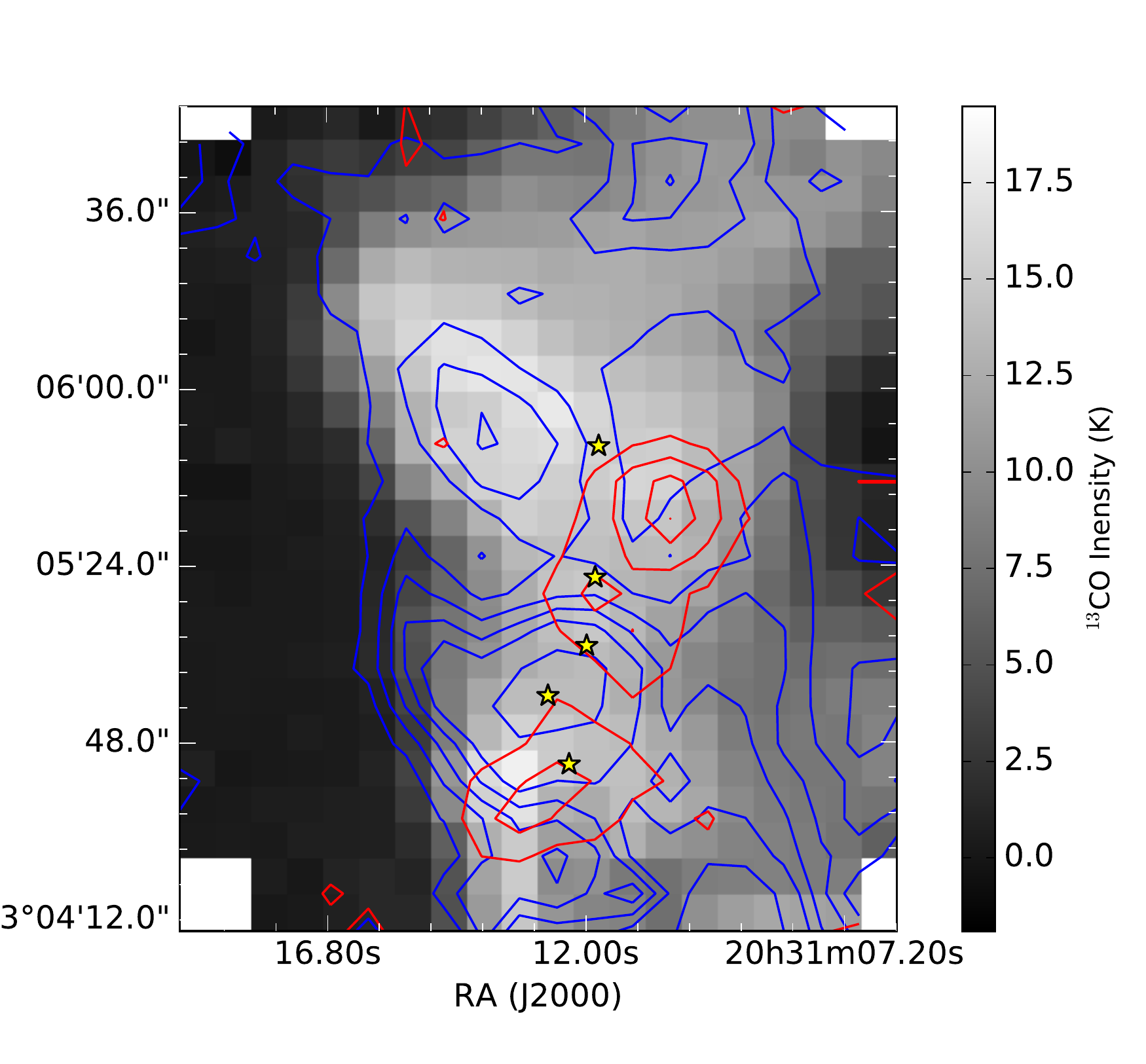}
    \caption{Integrated intensity of (a) $^{12}$CO and (b) $^{13}$CO line emission to highlight the spatial distribution of the molecular gas. Red and blue contours represent the red- and blue-shifted wings of the emission in the corresponding tracers, plotted over the background of total emission (gray-scale). Yellow stars indicate the locations of the protostars in the catalouge of \citet{protostars}. The gray-scale shows the integration over the entire spectral line, but the blue and red contour sets indicate emission over the velocity ranges of $v_{\mathrm{LSR}}=-15$ to $-5\mathrm{~km~s}^{-1}$ and $v_{\mathrm{LSR}}=0$ to $10\mathrm{~km~s}^{-1}$ respectively. \label{fig:intintmaps}}
\end{figure*}

\begin{figure*}
    \centering
    \includegraphics[width=0.45\textwidth]{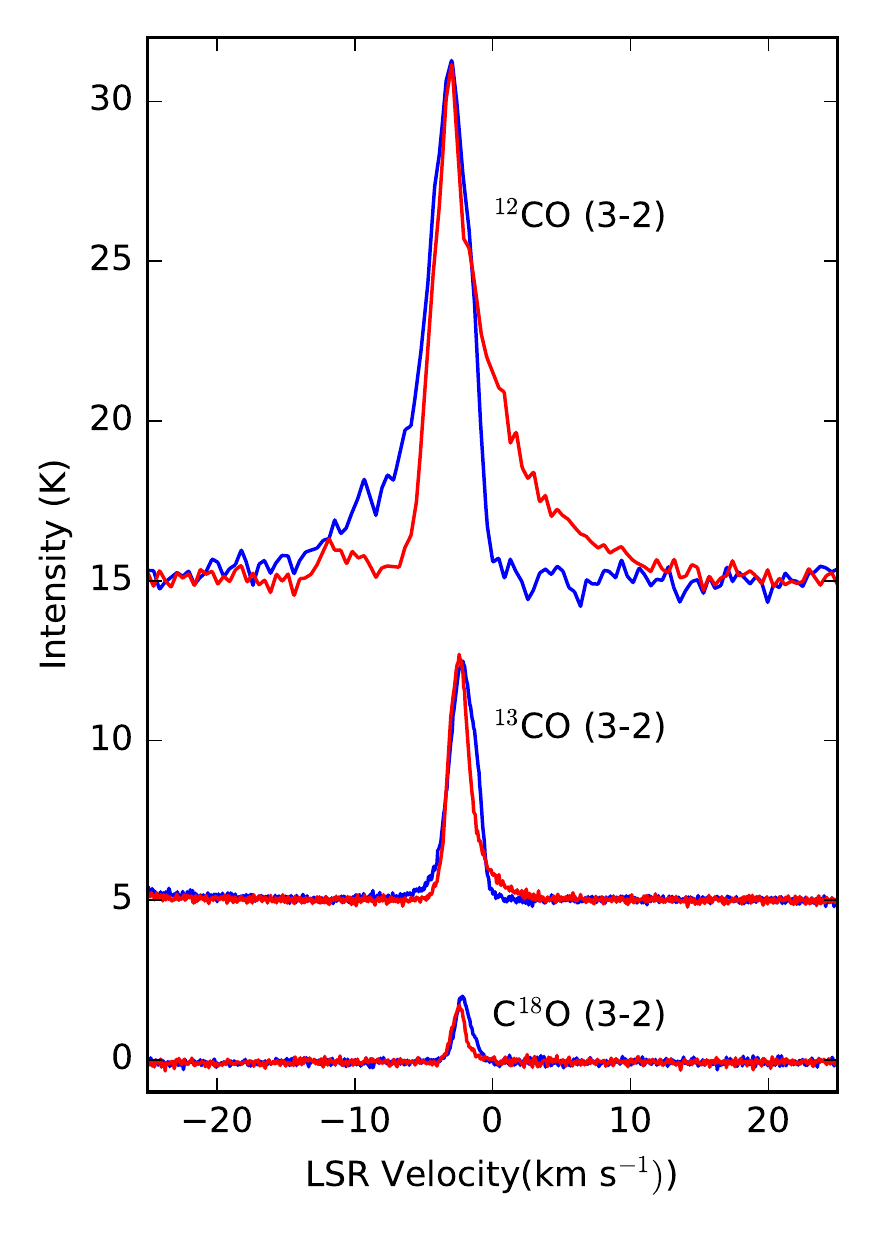}
    \includegraphics[width=0.45\textwidth]{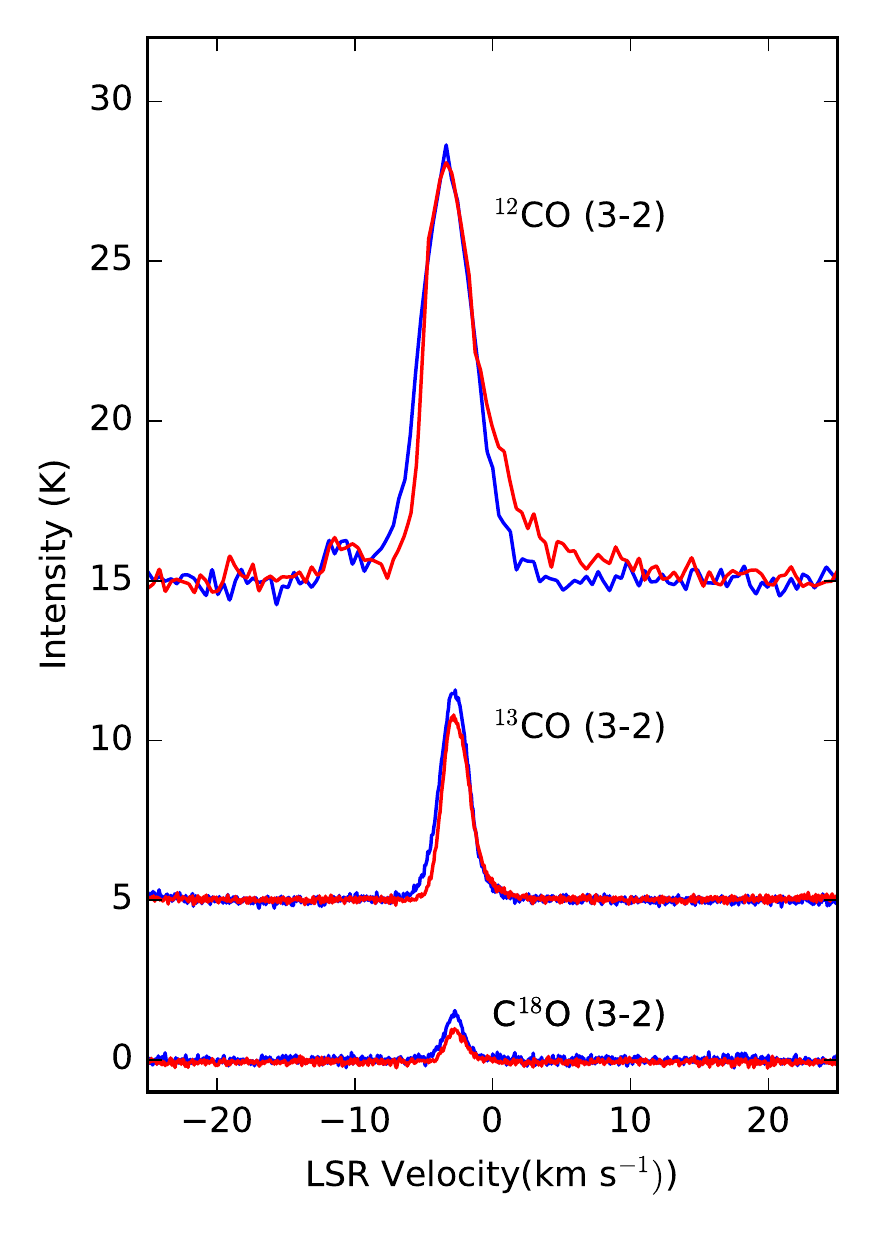}
    
    \caption{A side-by-side comparison between the outflows G81.435+2.147 (left) and G81.424+2.140 (right) in $^{12}$CO (offset +15~K), $^{13}$CO (offset +5~K), and C$^{18}$O lines. The asymmetry of the intensity around the central velocity indicates the strength of the outflow.  The red and blue colours denote spectra taken through the redshifted and blueshifted components of the outflow. The wing features that indicate the outflow are present in the high opacity $^{12}$CO line but the line profile is symmetric in the optically thin C$^{18}$O line.\label{fig:spectra}}
\end{figure*}

Figure \ref{fig:intintmaps} shows integrated intensity maps for the $^{12}$CO and $^{13}$CO emission. The contour sets highlight the two bipolar outflows.  The emission wings for the northern outflow are significantly brighter than the southern outflow.  Yellow stars indicate sources in the catalogue of \citet{protostars}. The northern outflow position is consistent with driving by J203111.98+430507.74 but the southern outflow is more confused. However, all of these sources have a positive spectral index indicating they are relatively young and embedded.

The bipolar outflows are clearly visible in the spectra shown in Figure \ref{fig:spectra}, which shows two spectra for each outflow. The red and blue spectra are sampled at positions containing the red- and blue-shifted lobes of the outflow respectively.  The line wings that indicate the outflow are clearly present in the high-opacity $^{12}$CO emission line, but the line profile becomes more symmetric in the lower-opacity $^{13}$CO and C$^{18}$O emission.  We will use these different opacities to estimate outflow properties in section \ref{sec:outflowprops}.

\begin{figure}
\centering
\includegraphics[width=\columnwidth]{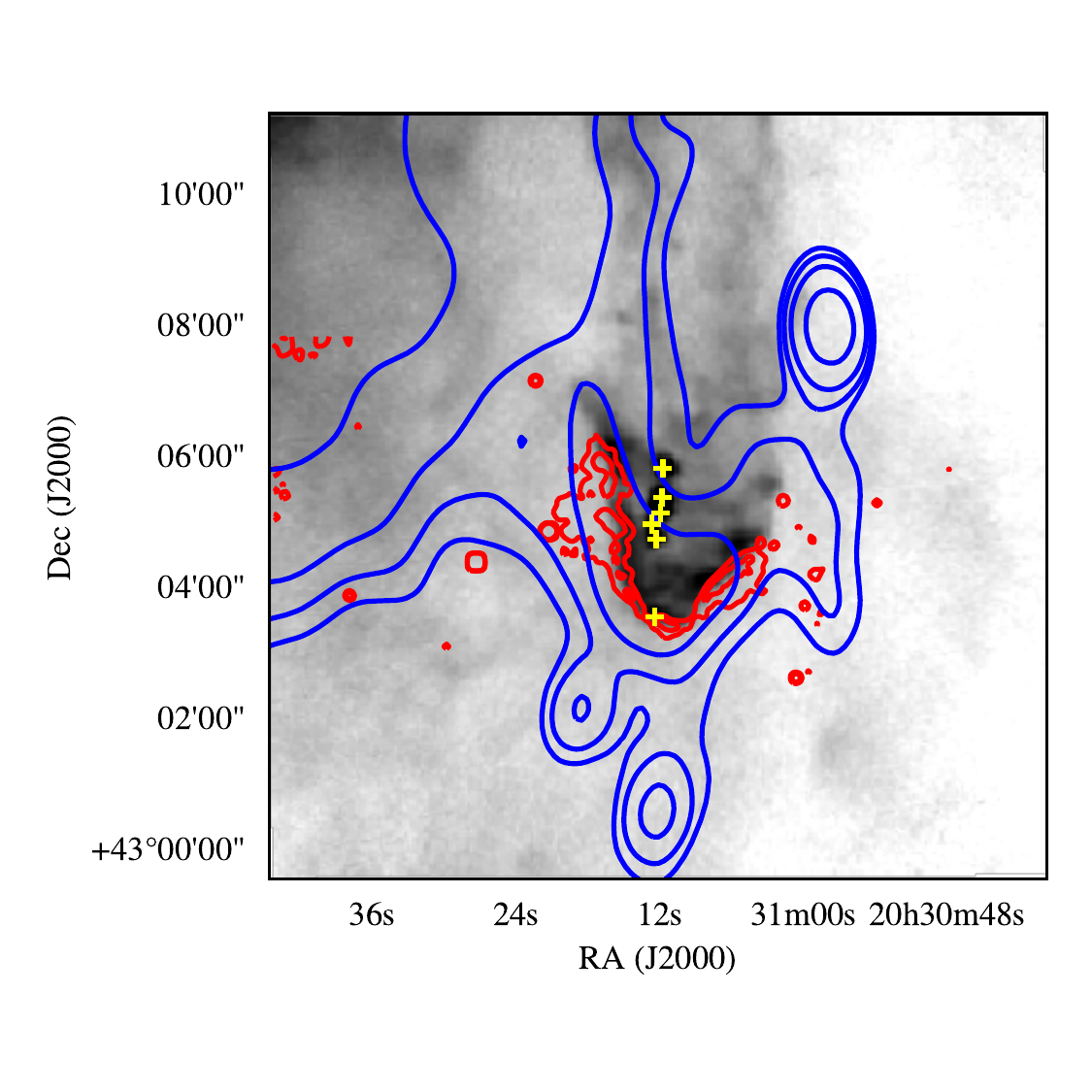}
\caption{Structure of the cometary feature seen in multiple wavebands.  The gray-scale image shows the PACS 70 $\mu$m data for the region on an arcsinh stretch, where darker colours indicate more intense emission.  The red contours show the H$\alpha$ flux with contour levels at 6 and $9 \times 10^{-17}\mbox{ erg s}^{-1}\mbox{ cm}^{-2}$ and the blue contours show the CGPS 21-cm radio continuum image with contours at 14.25, 14.75, 15.6 and 18 K.  The yellow $+$ signs indicate the six protostellar sources identified in \citet{protostars}. 
\label{fig:cometary_zoom}}
\end{figure}

We combine the CO emission line data with archival data from the dust continuum emission as observed by {\it Herschel}\footnote{Herschel is an ESA space observatory with science instruments provided by European-led Principal Investigator consortia and with important participation from NASA.} as part of the Hi-GAL program \citep[70 to 500 $\mu$m; OBSIDs 1342244170, 1342244171;][]{higal}, H$\alpha$ emission from the Isaac Newton Telescope Photometric H$\alpha$ Survey \citep[IPHAS;][]{IPHAS}, and the 21-cm radio continuum from the Canadian Galactic Plane Survey \citep[CGPS;][]{cgps}.  The infrared and radio data were obtained as fully calibrated images from their respective archives.  We process the IPHAS data to produce H$\alpha$ image by subtracting off a scaled $R$-band image to remove the stellar continuum.  We adopt the photometric calibration provided in the IPHAS metadata.  Figure \ref{fig:cometary_zoom} shows the cometary feature as seen in these archival data.  The infrared data highlight the column density of the neutral medium and reveal a limb-brightened cometary feature in the 70 $\mu$m data.  The H$\alpha$ and radio continuum data show emission from the ionized gas at the boundary of the cometary feature.  The comet head is directed south towards Cygnus OB2 and with the ``tail'' pointed directly away from the OB2 association.


\section{Results}
\label{sec:results}

\subsection{Mass of the cometary feature}

\begin{figure*}
    \centering
\includegraphics[width=0.45\textwidth]{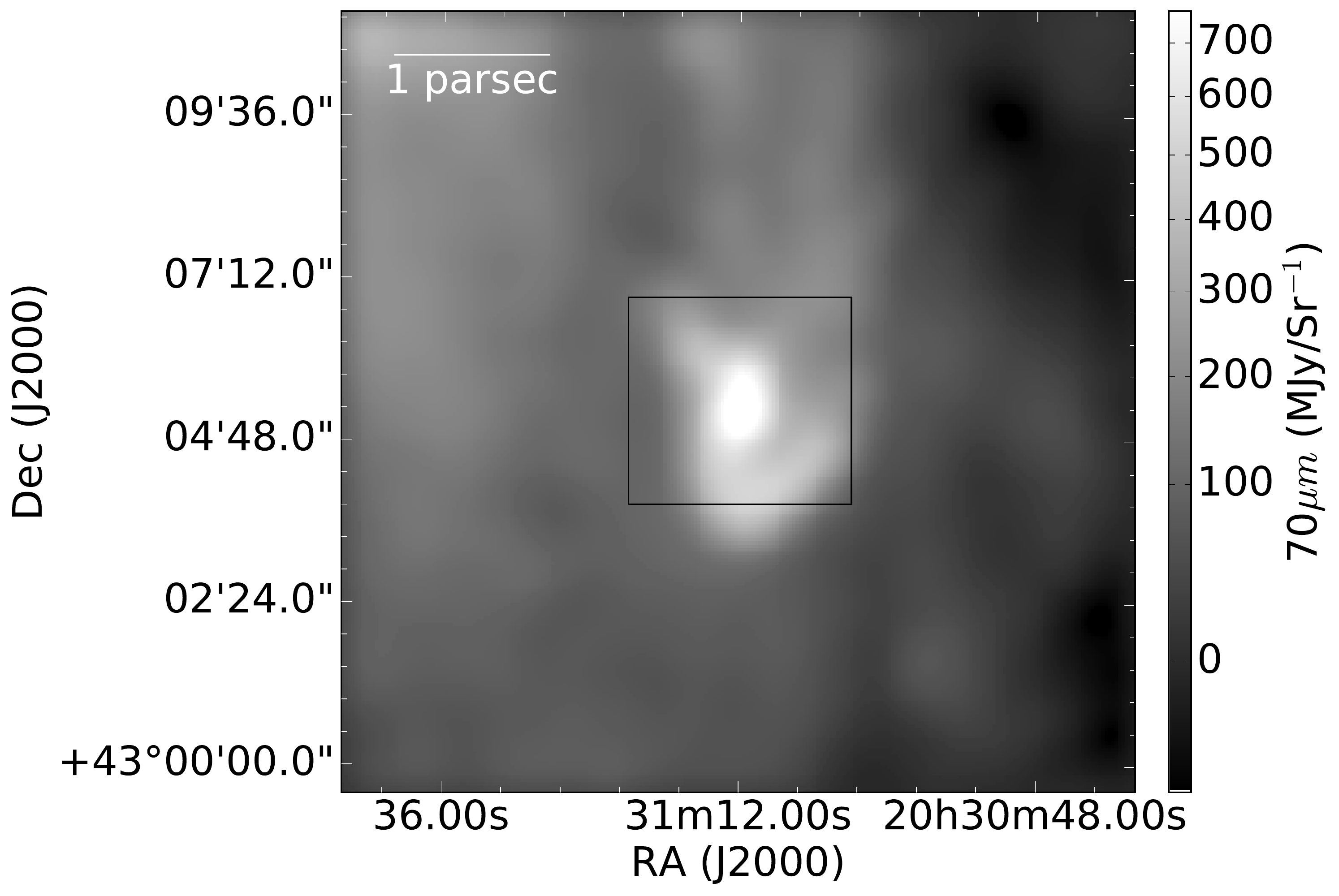}
\includegraphics[width=0.45\textwidth]{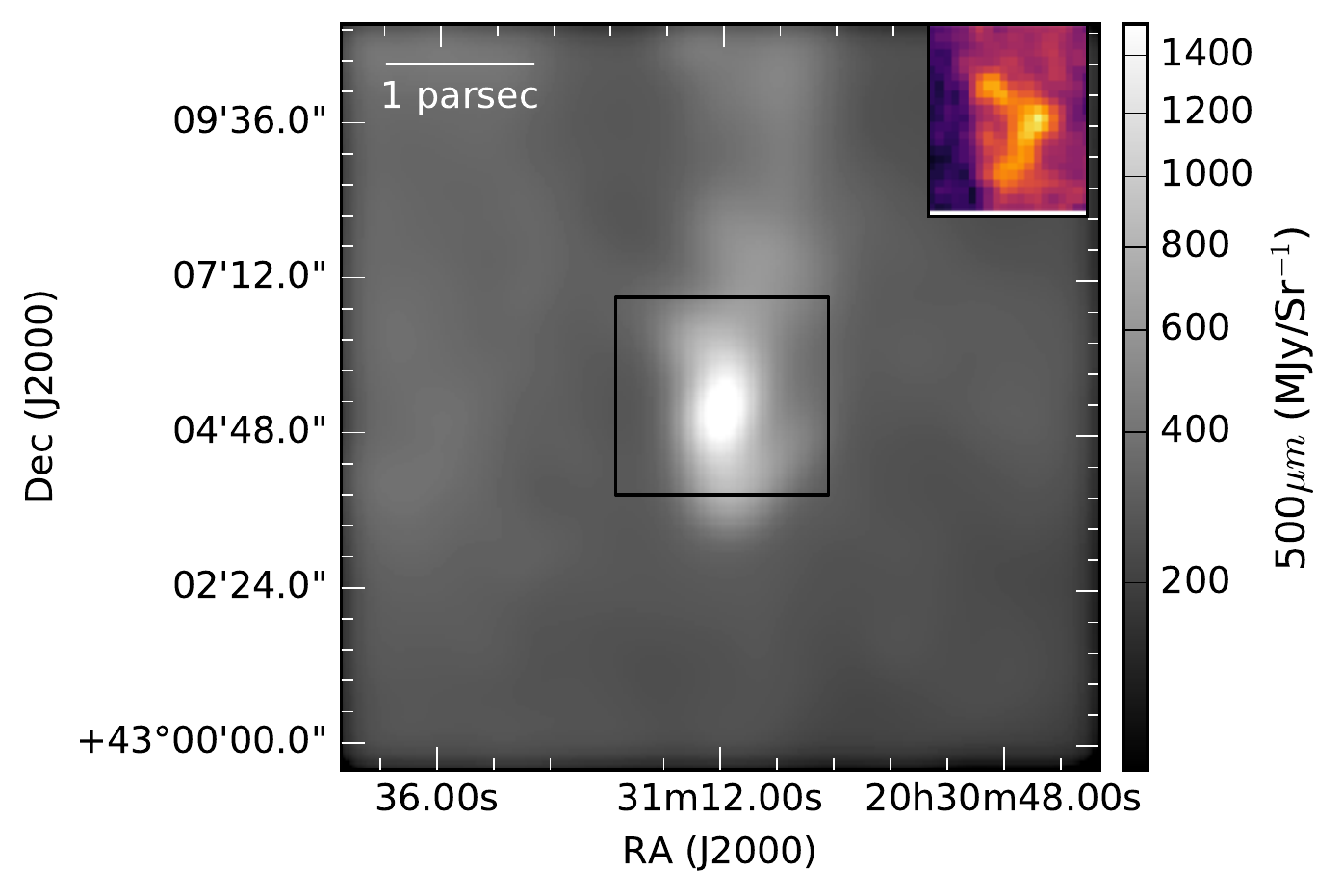}

    \caption{Far infrared imaging of the cometary feature from {\it Herschel} data showing the distribution of matter in the feature for wavelengths of 70 and 500 $\mu$m emission, in the left and right panels respectively.  The data are convolved to the resolution of the 500 $\mu$m data. The square in each figure shows the extent of the CO line mapping.  The upper right corner of the right panel shows the integrated intensity image for the $^{13}$CO data.  The figure shows the overall structure of the cometary feature, highlighting the enhanced short wavelength emission at the ``head'' of the comet and the tail of cooler, low column density gas behind the feature.\label{fig:irmaps}}
\end{figure*}

We measure the mass of the cometary feature using dust emission.  Specifically, we use the {\it Herschel} SPIRE (250, 350, and 500 $\mu$m) and PACS (70 and 160 $\mu$m) observations  of the region \citep{higal}, which we obtained from the Herschel archives.  Since the calibrated data were at different resolutions, we used a Gaussian kernel to convolve the short-wavelength data to match the $37''$ resolution of the 500 $\mu$m data (Figure \ref{fig:irmaps}).

Assuming the dust emission is optically thin, we model the emission from each band as 
\begin{equation}
    I_\nu=\Sigma_{\text{gas}}\kappa_{\nu}B_\nu(T)
    \label{eq:dustsed}
\end{equation}
    
where $B_\nu$ is the Planck function, $\Sigma_{\mathrm{dust}}$ is the surface density derived from dust and $\kappa_\nu$ is the opacity of the dust grains per unit gas mass.  We assume $\kappa$ is constant along the line of sight.  We adopt $\kappa_\nu$ from \citep{Hil}, 
\begin{equation}
     \kappa_{\nu}=0.1 \mathrm{\frac{cm^2}{g}}\left(\frac{\nu}{1\mathrm{~THz}}\right)^{\beta},
     \label{eq:dustsd}
\end{equation}
which includes a gas-to-dust ratio of 100 by mass to convert to an effective gas surface density and we assume $\beta=2$.



\begin{figure*}
    \centering
\includegraphics[width=0.45\textwidth]{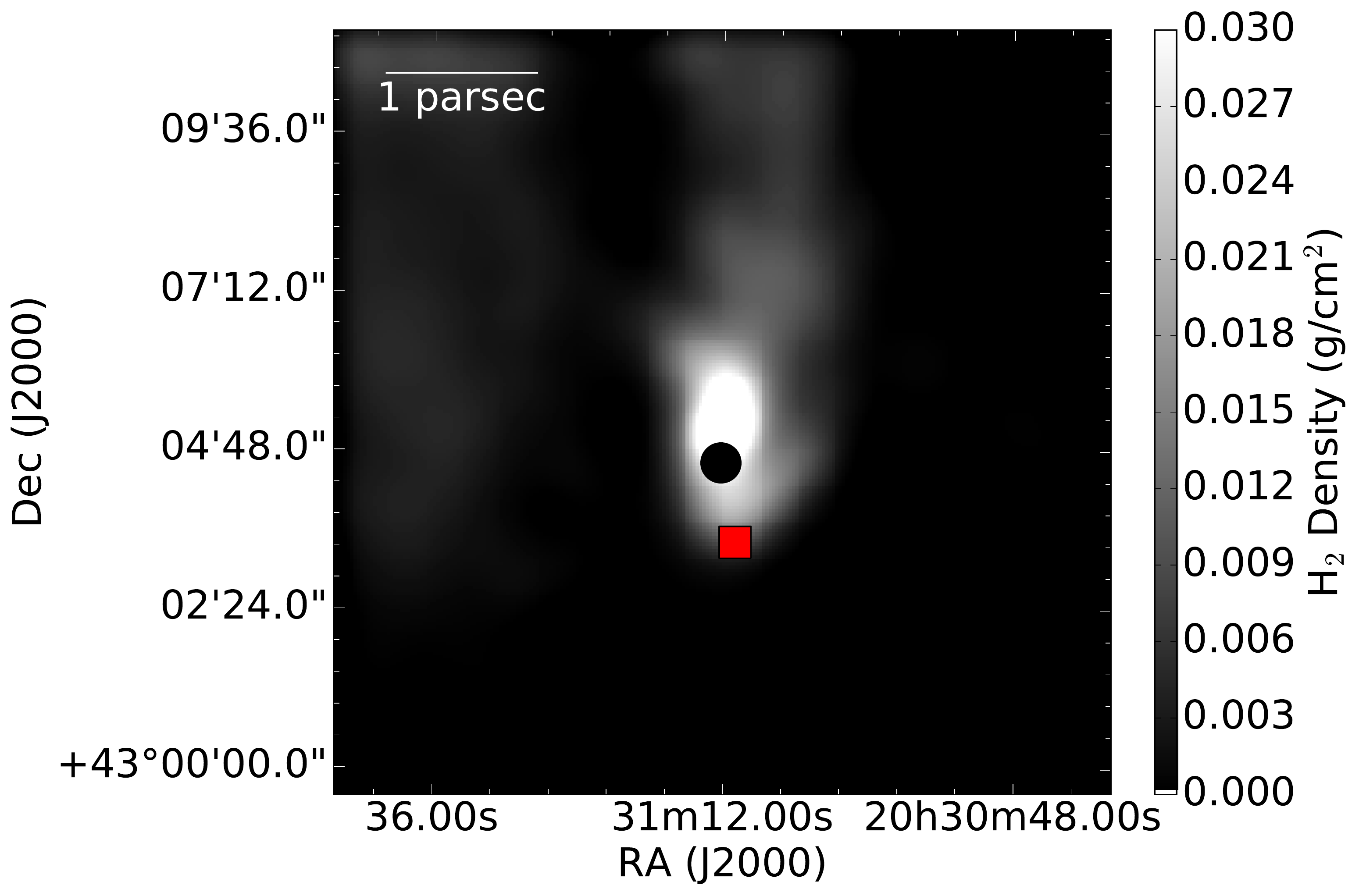}
\includegraphics[width=0.45\textwidth]{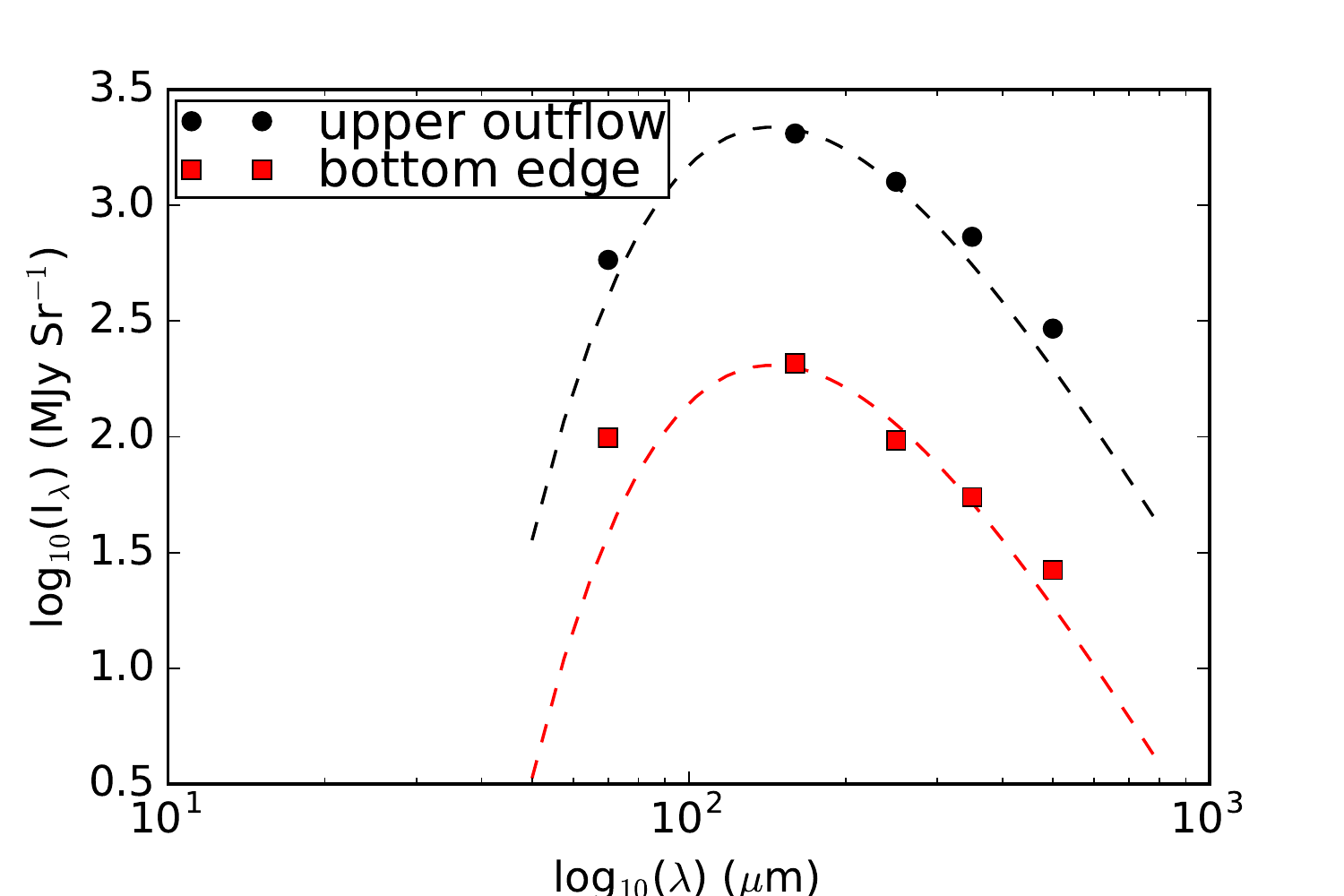}
\caption{(left) Surface density map of dust emission of the cometary region with background emission removed (right) Spectral Energy Distribution: fitted (dashed lines) and observed (squares/circles) for the second outflow source (20$^{\rm h}$31$^{\rm m}$12.31$^{\rm s}$, +43$^{\circ}$04$^{'}$43.61$^{''}$) and the bottom edge of the cometary feature (20$^{\rm h}$31$^{\rm m}$12.19$^{\rm s}$, +43$^{\circ}$03$^{'}$07.25$^{''}$). Positions are shown in black circles and red squares respectively in the left panel. The SED at the bottom edge shows that the 70~$\mu$m data sits above the fit for the rest of the curves, indicating the presence of a second hot dust component. \label{fig:seds}}  
\end{figure*}

We have two parameters $\Sigma$ and $T$ in our model, which we estimate by fitting the model (Equation \ref{eq:dustsed}) to each position in the data.  Figure \ref{fig:irmaps} shows that the region contains spatially varying background emission.  We define a background level from a region near the cometary feature in each image and subtracting the average in that region from the data.  On examining the spectral energy distributions of different positions (e.g., Figure \ref{fig:seds}), we find that the 70 $\mu$m data are not well represented by a single dust temperature model, with the bright 70 $\mu$m emission found at the limb of the comet (see also the full resolution image in Figure \ref{fig:cometary_zoom}).  We thus omitted the 70 $\mu$m data from the fit and adopted a fixed dust temperature of $T=20$~K, which should trace the bulk of the mass in the star forming comet.  Under these assumptions, we derive the column density map (Figure \ref{fig:seds}) and find the mass of the feature to be $M=110~\mathrm{M}_\odot$, assuming a distance of 1.4 kpc.

\subsection{CO line Emission: Optical Depth and Column Density}
\label{sec:abund}
We also derive the column density of the CO isotopologues using the multi-line observations.  Specifically, we assume local thermodynamic equilibrium so the spectral line emission is modelled as \citep{mangum}:
\begin{equation}
T_\text{R}=f[J_\nu (T_{\text{ex}})-J_\nu(T_{\text{bg}})](1-e^{-\tau_{\nu}}).\label{eq:tbmodel}
\end{equation}
Here $T_\text{R}$ is the observed radiation temperature, $T_\text{ex}$ is the molecular excitation temperature, $T_{\text{bg}}$ is taken as the cosmic microwave background temperature ($\approx$ 2.73 K), $J_\nu (T)=(h\nu/k)/(\exp[h\nu/(kT)]-1)$ is the compensated radiation temperature, and $\tau$ is the optical depth. The factor $f$ corresponds to the beam dilution factor, which we assume is 1 (i.e., the emission is beam-filling).



As $^{12}$CO is a much more abundant species in molecular clouds, compared to $^{13}$CO and C$^{18}$O, we assume the emission is optically thick in $^{12}$CO emission, particularly at the line centre. In that case, 
\begin{equation}
J_\nu (T_{\text{ex}})-J_\nu(T_{\text{bg}})\approx T_\mathrm{max},
\end{equation}
where $T_\mathrm{max}$ is the maximum radiation temperature of the $^{12}$CO emission along each line of sight.  In that case, the molecular excitation temperature becomes
\begin{equation}
    T_{\text{ex}}=\frac{h\nu/k}{\ln\left[1+\frac{h\nu/k}{T_\mathrm{max}+J_\nu(T_{\text{bg}})}\right]}
\end{equation}
We assume the $^{13}$CO and C$^{18}$O have the same excitation temperature as $^{12}$CO, and that the C$^{18}$O emission is optically thin. We can then express the column density for the top state of the C$^{18}$O transition as:
\begin{eqnarray}
    N_u& =&\frac{8\pi \nu_0^3}{c^3 A_{ul}}\frac{1}{e^{\frac{h\nu_0}{kT_{\text{ex}}}}-1}\int \tau_\nu dv, \nonumber\\
& = & \frac{8\pi \nu_0^3}{c^3 A_{ul}}\frac{1}{e^{\frac{h\nu_0}{kT_{\text{ex}}}}-1} \frac{\int T_\text{R} dv}{J_\nu (T_{\rm ex})-J_\nu(T_{\text{bg}})}.
\end{eqnarray}
Here, $\nu_0$ is the equivalent frequency and $A_{ul}$ is the Einstein coefficient for $u=3$ to $l=2$ transition. We extrapolate to the total column density of the species using the partition function ($Q$), which is well approximated as \begin{equation}
Q\approx \frac{kT}{hB_0}\exp\left(\frac{hB_0}{3kT}\right). 
\end{equation}
With these assumptions, the total column density is 
\begin{equation}
    N_{\text{tot}}=\frac{Q}{g_u}\exp\left(\frac{E_u}{kT_{\text{ex}}}\right)N_u. \label{eq:Ntot}
\end{equation}
For this calculation, the Einstein coefficient $A_{ul}=6.011\times 10^{-7}$ s$^{-1}$, $\nu_0=330.588$ GHz and the rotational constant $B_0=54891.42$ MHz, with values from the LAMDA\footnote{\url{http://home.strw.leidenuniv.nl/~moldata/}} \citep{lamda} and NIST\footnote{\url{ https://physics.nist.gov/PhysRefData/MolSpec/}} databases. \\

The $^{13}$CO line is not necessarily optically thin.  In this case, Equation \ref{eq:tbmodel} implies 
\begin{equation}
        \tau_{\nu}=-{\rm ln}\left[1-\frac{T_{\rm R}}{J(T_{\rm ex})-J(T_{\rm bg})}\right]
\end{equation}
We compute a total column density of $^{13}$CO (Equation \ref{eq:Ntot}) using $A_{ul}=6.038\times 10^{-7}$ s$^{-1}$, $\nu_0=330.588$ GHz and $B_0=55101.01$ MHz.

\begin{figure}
    \centering
    \includegraphics[width=\columnwidth]{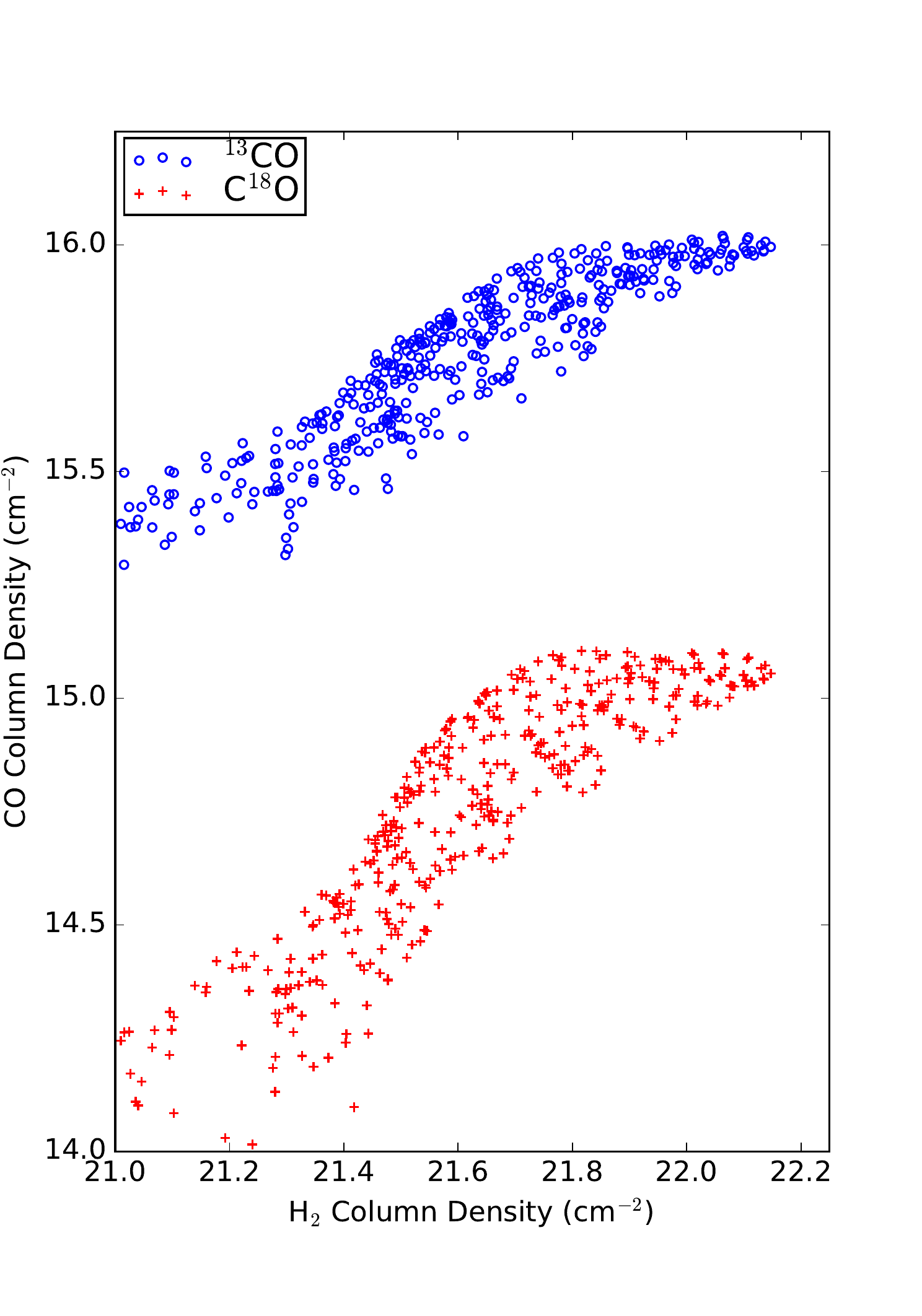}
    \caption{CO isotopologue column densities relative to H$_2$ column density estimate obtained from dust continuum emission.  Both species show typical fractional abundances, and some signs of depletion onto dust grains at the highest dust column densities.  This figure shows our estimates of the different column densities are broadly consistent with expectations for a typical cloud of star forming molecular gas.}
     \label{fig:coh2}
\end{figure}

Fig. \ref{fig:coh2} shows the correlation between the H$_2$ and the CO column densities.  We convolve the molecular line column density maps to match the $37''$ resolution of the dust map.  This figure shows typical fractional abundances of $10^{-6}$ and $1.5\times 10^{-7}$ for the $^{13}\mathrm{CO}$ and C$^{18}$O respectively, which are comparable to the canonical values \citep{wilson94}.  However, the $^{13}$CO emission shows a shallower increase with increasing column density compared to the C$^{18}$O.  At large values of the dust column density, both species appear to saturate indicating some depletion onto dust grains.

\subsection{Properties of the Outflows}
\label{sec:outflowprops}
We also estimate the mass, momentum and kinetic energy of the molecular outflows from the line emission data. In particular, we use $^{13}$CO column density data for this purpose.  Using Equation \ref{eq:Ntot} in the previous section, the column density of $^{13}$CO as a function of position and velocity is:
\begin{equation}
   N_{^{13}\rm CO} (x,y,v)=\frac{8\pi \nu_0^3Q_{\rm rot}}{7c^3 A_{ul}}\frac{e^{\frac{E_u}{kT_{\text{ex}}}}}{e^{\frac{h\nu_0}{kT_{\text{ex}}}}-1}\tau_\nu (\mathbf{x},v)\, \delta v.
\end{equation}
Here, $\mathbf{x}$ is the position along a position-velocity (PV) slice, running through the blue and red shifted region of an outflow. We have also used $\int \tau_\nu (\mathbf{x},v) dv \approx \tau_\nu (\mathbf{x},v)\, \delta v $ where $\delta v$ is the channel width, and $g_u=2J_u+1=7$ . Using our typical fractional abundance of $^{13}\mathrm{CO}$ derived from dust in section \ref{sec:abund}, the column density of H$_2$ is $N_{\rm H_2}(x,y,v)=10^6 N_{^{13}\rm CO} (x,y,v)$. This expression gives us an estimate of the mass as a function of position offset and velocity by assuming a mean particle mass of $\mu =2.4m_\mathrm{H}$, where we include the mass of helium.


The signal-to-noise ratio of the $^{13}$CO emission is low in the wings of the outflows, so we infer the brightness of the $^{13}$CO from the brighter $^{12}$CO data \citep{arce01}.  First, we estimate the line centre by fitting a Gaussian to the C$^{18}$O data extracted on the same PV slice. Then we fit a quadratic function to the $^{12}$CO/$^{13}$CO line ratio separately for blue and red-shifted regions.  Again following \cite{arce01}, we assume a minimum value of the line ratio is found in the optically thin case where the relative abundance of the species, assumed to be $65$ \citep{wilson94}, sets the ratio.

With these assumptions, we calculate the mass at every position and velocity in the outflow.  We calculate the total mass in the outflow by summing the contributions over the region, subtracting off the contribution associated with the main molecular cloud as defined by the C$^{18}$O.  We also calculate the projected momentum ($p \cos i = \sum M|v-v_0|$) and energy ($E\cos^2 i = \sum M(v-v_0)^2/2$) for the outflows and give the results in Table \ref{tab:outflows}.   We also list the position angle (PA) along which the the position-velocity slice is extracted, measured in degrees east of north. The $\cos i$ factor accounts for the inclination of the outflow with respect to the line of sight. For comparison, the two outflows are at the upper end of momentum and energy distributions compared to the population of outflows seen in the nearby Perseus molecular cloud \citep{arce10}, but the outflows in Cygnus X are not exceptional.

\begin{table}
    \centering
    \begin{tabular}{lrr}
    Property & Outflow 1 & Outflow 2 \\
         \hline
         RA (J2000)  & 20:31:12.5 & 20:31:12.3		\\
         Dec (J2000) &+43:05:42 &+43:04:53  \\
         PA ($^\circ$) &48.01  &141.34 \\
         $M_\mathrm{o}~(\mathrm{M}_\odot)$ &12 & 3.6 \\
         $p_\mathrm{o} \cos i~(\mathrm{M}_\odot~\mbox{ km s}^{-1})$ &26  & 8.5\\
         $E_\mathrm{o} \cos^2 i ~(10^{44}\mbox{ erg})$ & 8.8 & 2.5\\
         \hline
    \end{tabular}
    \caption{Properties of molecular outflows found in the cometary feature.}
    \label{tab:outflows}
\end{table}

\section{Discussion}
\label{sec:discuss}

Here we discuss the properties of the star forming molecular feature in the context of the broader star forming environment of the Cygnus X region.  

\subsection{Ionizing Sources}
\label{sec:ionizing}

The cometary feature shows a rim of radio continuum emission (Fig.~\ref{fig:cometary_zoom}) visible at 1420~MHz in the Canadian Galactic Plane Survey \citep[CGPS,][]{cgps} and the NRAO VLA Sky Survey \citep[NVSS,][]{nvss}, outlining the CO and~IR emission of the cometary feature. Assuming this is free-free emission from the hot ionized gas at the edge of the cometary feature, we can calculate the ionizing flux, required to maintain this thermal emission.

The radio continuum emission is only marginally resolved, so we calculate the plasma properties of the region using the flux density of emission and assuming a geometry based on the H$\alpha$ emission. The total integrated flux density $S_\nu$ at a frequency of $\nu = 1420$~MHz results in $60 \pm 6$~mJy. Using the expression for the brightness temperature from \citet{rohlfs}, the flux density for a source of optically-thin, free-free emission is:
\begin{equation}
S_\nu = 2.553\mbox{ mJy} \left(\frac{E_V}{\mathrm{pc^3~cm^{-6}}}\right) \left(\frac{d}{\mathrm{kpc}}\right)^{-2} \left(\frac{T_e}{\mathrm{K}}\right)^{-0.35} \left(\frac{\nu}{\mathrm{GHz}}\right)^{-0.1},
\end{equation}
where $T_e$ is the electron temperature, $d$ is the distance to the source, and $E_V$ is the volume emissivity:
\begin{equation}
    E_V = \int n_e^2 dV.
\end{equation}
For an electron temperature of $8000 \pm 2000$~K \citep[e.g.,][]{osterbrock} and a distance of 1.4~kpc to the cometary feature we calculate a volume emissivity of 
\begin{equation}
   E_V = 1100\pm 200~~{\rm pc^3\,cm^{-6}}.
\end{equation}


Since the radio data does not fully resolve the rim of thermal emission, we use the IPHAS image to determine the location of the ionized gas. The resolution of the CGPS images is about $1'$
and of the NVSS images about $45''$, while the rim of ionized gas is about $\Delta r = 5''$ to $10''$ wide when viewed in the IPHAS data (see Fig.~\ref{fig:cometary_zoom}).  We model the region as a thin, hemispherical shell with radius of $r = 1.5' = 0.6$~pc.  We then calculate the typical electron density as $n_e \approx [E_V/(2\pi r^2 \Delta r)]^{1/2} = 110 \pm 10 \mbox{ cm}^{-3}$.  We also estimate the mass of ionized gas as $M = m_\mathrm{H} n_e 2\pi r^2 \Delta r \sim 0.2~M_\odot$.

We also calculate the required ionizing flux using the IPHAS image of the region \citep{IPHAS}. To make this estimate, we measure the average brightness of the cometary feature in a continuum subtracted H$\alpha$ image finding $\langle f \rangle = 2\times 10^{-16}\mbox{ erg s}^{-1}\mbox{ cm}^{-2}$ coming from each 0.33 arcsec pixel, which implies an emergent intensity of $3\times 10^{-5}\mbox{ erg s}^{-1}\mbox{ cm}^{-2}\mbox{ sr}^{-1}$.  Assuming no extinction, this measurement in turn implies an emission measure of $\mbox{EM} = 1000 \mbox{ pc cm}^{-6}$ using the calibration of \citet{haffner98}, again taking $T_e = 8000$~K.  Using our model as before, the path length is determined by the distance through the shell, where the typical path length is $\gtrsim 2 \Delta r$. In this case, the typical electron density is $n_e \approx [\mathrm{EM}/(2\Delta r)]^{1/2} \approx 100\pm 20 \mbox{ cm}^{-3}$, consistent with the estimate from the radio continuum emission.

We explore what sources could be providing ionizing flux that maintains the thermal emission by setting the recombination rate for the region equal to the geometrically diluted ionizing photon flux for a source that is a distance $d$ away.  Expressing this condition in terms of the Case B recombination coefficient $\alpha_B$, the emission measure and the ionizing photons produced per second ($Q_0$) yields the steady-state requirement that:
\begin{equation}
\mathrm{EM}\,\alpha_B = \frac{Q_0}{4\pi d^2} \equiv q_0.
\end{equation}
Taking $\alpha_B \approx 2.56\times 10^{-13} T_4^{-0.83} \mbox{ cm}^{3}\mbox{ s}^{-1}$, where $T_4 = (T_e/10^4\mbox{ K})$  gives $q_0= 1\times 10^9 \mbox{ cm}^{-2}\mbox{ s}^{-1}$ for $T_4 = 0.8$ \citep{Drn}.

\begin{figure}
    \centering
    \includegraphics[width=\columnwidth]{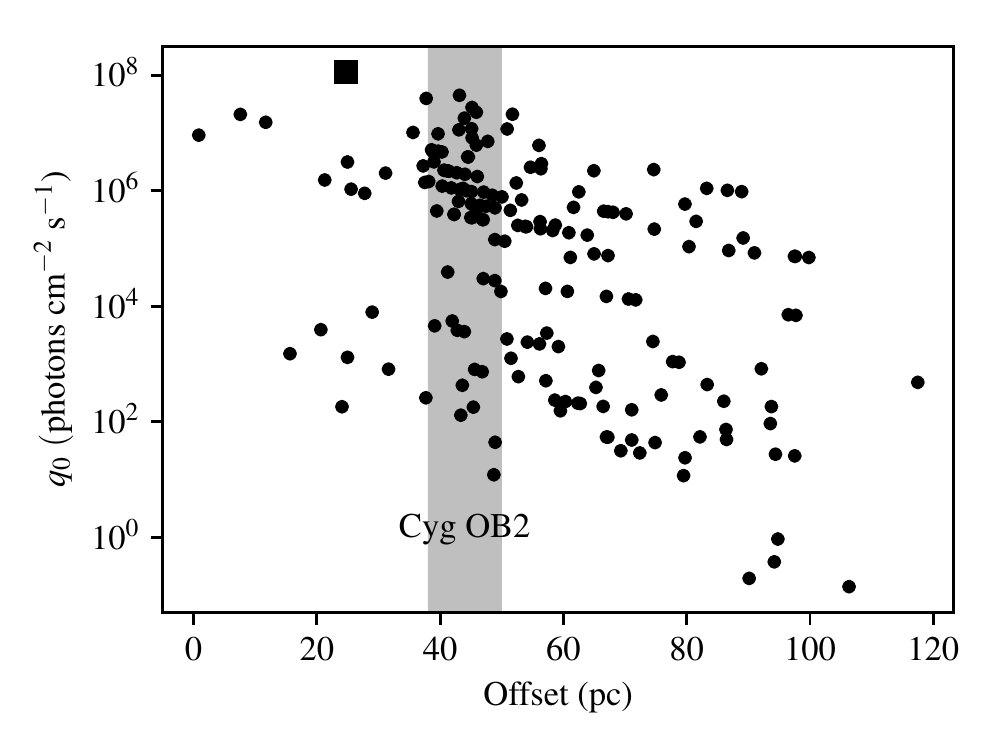}
    \caption{Maximum flux of ionizing photons at the location of the cometary feature based on the catalog of \citet{B1}. The largest contribution to the modelled flux from an individual star arises from BD+43~3654, indicated with a filled square. The dominant fraction of the emission comes from the collective action of the Cyg OB2 complex.}
    \label{fig:obflux}
\end{figure}

We evaluate the possible ionizing sources that could be powering this feature.  We explore the catalog of \citet{B1}, searching for O and B stars near the cometary feature.  For each star in the catalog, we calculate the number of ionizing photons produced using the model atmospheres of \citet{sternberg03}.  For stars with spectral types later than B0, we estimate the ionizing photons by extrapolating the $T_{\mathrm{eff}}$ vs $Q_0$ curves to lower temperatures.  However, these stars produce so few ionizing photons that they could not contribute significantly to the ionization.  For each star, we calculate the maximum possible flux at the location of the cometary feature by assuming the stars are at the same distance as the feature and there is no extinction between the star and the cometary feature.  We plot the results of this analysis in Fig. \ref{fig:obflux}.  The total ionizing flux provided by all cataloged stars is $q_0 = 5\times 10^8\mbox{ photons cm}^{-2}\mbox{ s}^{-1}$, close to the required budget based on the emission measure.  In this simple model, the star BD+43 3654, an O4If star at a projected offset of 24 pc provides 22\% of the total flux at the source.  However, BD+43~3654 is located to the north of the cometary feature, and the H$\alpha$ and radio geometry imply illumination from the south.  Thus, BD+43~3654 is unlikely to be the primary ionizing source.  It cannot be at the same projected distance as the cometary feature or else the ionization limb would not have the orientation that it does.  Other, nearer stars also have unfavourable geometries.  In contrast, the collective action of stars in Cyg OB2 provide $>50\%$ of the total possible photons in our model and are oriented in the correct direction to shape the cometary feature.   Given the number of ionizing photons is of order the amount provided by Cyg OB2, we regard the cometary feature as likely being photoevaporated by radiation from the Cyg OB2.

We attribute the lack of a perfect agreement between the ionizing photon budget required to maintain the emission and that provided by the known stars to the incompleteness of the OB star catalogues in this regions.  Many sources of ionizing photons could be obscured by dust screens from our perspective, rendering the catalogues incomplete.

\subsection{Internal conditions of the cometary feature}

While frequently correlated, the causal connection between cometary features and the triggering of star formation by  feedback from local massive stars is not clear \citep{Dale}, particularly for gravitationally bound cloud fragments. The star formation could be proceeding because of spontaneous gravitational collapse or compression caused by nearby supernova blast waves. To assess the likelihood of the star formation being triggered, we must look at the dynamical quantities in the cometary feature. In particular, the feedback-driven implosion of the globule requires that the thermal pressure in the surrounding {\sc Hii} rim and the external radiation pressure due to the ionizing flux from OB2 complex to dominate the gravitational energy of the cometary feature.

We parameterize the different physical effects in terms of the energy densities and pressures to evaluate what effects govern the evolution of the feature. For global quantities we convert energies to energy densities by assuming a spherical geometry for the object.  Taking the distance to be 1.4 kpc, we estimate the radius as $R \approx$ 0.31 pc. The gravitational potential energy density, assuming uniform density is $u_\mathrm{grav} = 9GM^2 / (20 \pi k_\mathrm{B} R^4) \approx 4.0\times10^6$~K~cm$^{-3}$. We also calculate the internal pressure from the velocity dispersion in the molecular gas.  We estimate the line width from C$^{18}$O (3-2) data by fitting a Gaussian profile to the mean spectrum over the mapped region, finding $\sigma_v \approx$ 0.68~km~s$^{-1}$.  We use the C$^{18}$O since the line is the most optically thin tracer of the molecular gas and is unaffected by the line broadening due to outflows.  Then the internal gas pressure is $P_\mathrm{int}/k_\mathrm{B} = \rho \sigma_v^2/k_B \approx 1.9 \times 10^{6}~\mbox{K cm}^{-3}$.  

For thermal and radiation pressures we use the values calculated in Section \ref{sec:ionizing}.  We obtain thermal pressure in the ionized gas as $2 n_\mathrm{e} T_\mathrm{e} \approx 2.24 \times 10^6$ K cm$^{-3}$ and radiation pressure as  $P_\mathrm{rad}=h\nu q_0/(c k_\mathrm{B}) \approx 1.25\times 10^4$~K~cm$^{-3}$, where $T_\mathrm{e}$ and $n_\mathrm{e}$ are the electron temperature and mean electron number density in the ionized rim, $q_0$ is the incident ionizing flux density.  We also estimate the contributions to the energetics of the cloud by taking the energy for the molecular outflows (Table \ref{tab:outflows}) and dividing by the volume of the object.  This represents an upper limit since the molecular outflows are not likely to contribute all this energy to the support of the region. The outflow wing velocities ($v_\mathrm{wing}\sim 5\mbox{ km s}^{-1}$) are significantly larger than the escape speed for the cometary feature ($v_\mathrm{esc} = 1.2~\mbox{km s}^{-1}$).

Considering these relative energy densities and pressures, the cometary feature is gravitationally bound with kinetic energies comparable to the binding energy, specifically $K = 0.75 U_g$.  However, the external pressure is comparable to the internal pressure for the feature, which will encourage gravitational collapse.  The line widths of the feature are larger than expected from a virialized object and could be indicative instead of free-fall collapse or radiative driven implosion \citep{lefloch94}.

For full consideration of the energetics, we write the virial theorem for the cometary feature,
\begin{equation}
    \frac{1}{2}\ddot{I}  = U_g + 2 K - \oint_A r P dA +  \frac{1}{8\pi} \int (B^2 - B_0^2) dV,
\end{equation}
where the terms $U_g=- 3GM^2/(5R)$ and $K= 3 M \sigma_v^2/2$ represent the potential and kinetic energies of the cometary feature, the values of which can be found in  Table \ref{tab:compare}, as $U_g=-2\times 10^{45}$ erg, $2K=3\times 10^{45}$ erg. The third term is the mean external pressure at the virial surface. Assuming a spherical geometry, we calculate $\oint_A r P dA = 1.7\times 10^{45}$ erg over $2\pi$~sr representing the hemisphere bounded by the ionizing gas.  If the magnetic field term (i.e., the last term) in this equation is zero, then $\ddot I<0$ and the feature will collapse on a free-fall time.

The last term represents the unknown total magnetic energy of the cometary feature ($B$) minus that due to the background field ($B_0$). Given the other physical effects that are governing the energetics of this feature, we calculate the field strength required to achieve virial balance  ($\ddot{I}=0$) as 70~$\mu$G, taking $B^2 \gg B_0^2=(5~\mu \mathrm{G})^2$ \citep{crutcher}. A high magnetic field strength in a collapsing region favours the possibility of external influence such as photoevaporation. While reasonably strong, fields of this strength or larger are seen in cometary features \citep{pattle18}, where observations suggest a value between 170 and 320 $\mu$G with field lines aligned with the photoevaporative flow. Whether the field  can support the feature under discussion against collapse depends on its geometry and strength, neither of which can be further assessed without dedicated observations.




 
\bigskip
 


\begin{table*}
    \centering
    \begin{tabular}{lcrrrr}
    Property & Expression & & Cometary Feature & &Eagle Nebula Pillar II  \\
         \hline
         Mass (M$_\odot$) & $M$ & & 110 & &250 \\
         Size (pc) & $R$ & &0.31 & &0.15 \\
         Electron density in ionized rim (cm$^{-3}$) & $n_e$ & &140 & &4000  \\
         Potential energy density (K cm$^{-3}$) & $9GM^2 / (20 \pi k_\mathrm{B} R^4)$ &  &$3.8\times 10^6$ & &3.7$\times 10^8$  \\
         Internal pressure (K cm$^{-3}$)& $\rho \sigma_v^2/k_B$& &$1.9\times 10^6$& &1.2$\times 10^8$  \\
         Incoming ionizing flux (cm$^{-2}~{\rm s}^{-1}$) &$q_0$ & &$2.1\times 10^9$ & &4.6$\times 10^{11}$ \\ 
         Radiation pressure (K cm$^{-3}$)&$h\nu q_0/(c k_\mathrm{B})$ & &$1.2 \times 10^4$ & & 2.7$\times 10^6$  \\
         Thermal pressure at the tip (K cm$^{-3}$)&$2 n_\mathrm{e} T$ & &$2.2\times 10^6$ & &6$\times 10^7$ \\
         Outflow energy density (K cm$^{-3}$) &$E_\mathrm{o} \cos^2 i / (\frac{4}{3}\pi R^3)$ & &$<1.7\times 10^6$ & &$\cdots$\\
         Binding energy (10$^{45}$ erg) & $3GM^2/5R$ & &2 & &21  \\
         Photoevaporation mass-loss rate (M$_\odot$ Myr$^{-1}$) & $\dot{M}$& & 15  & &70\\
         Photoevaporation lifetime (Myr) & $M/\dot{M}$ & & 7 & &3\\
         \hline
    \end{tabular}
    \caption{Comparison of dynamical properties between the cometary feature and Eagle Nebula}
    \label{tab:compare}
\end{table*}
\subsection{Timescales for Evolution of the Cometary Feature}
While Cygnus OB2 is the likely candidate for illuminating the limb of the cometary feature, the process of star formation is currently ongoing inside the cold cloud.  To evaluate whether Cyg OB2 could have triggered the formation of stars inside the region, we examine the timescales for star formation to be triggered within the cometary feature.  We consider a triggering scenario where Cygnus OB2 forms, creates an ionization front in a neutral cloud that then propagates to the location of the cometary feature, prompting an increase in the star formation in this region.

{\it Source of Triggering}: The study of \citet{wright15} argues that massive star formation occurred in Cyg OB2 continuously between 1 and 7 Myr ago, so that the region has been producing significant amounts of ionizing radiation. Considering only the stars in the \citet{B1} catalog found in the central Cyg OB2 region, the current production of ionizing photons is $10^{50.8}\mbox{ s}^{-1}$, and the photon rate over the past few Myr is likely declining as the stellar population ages. Therefore, this value should be considered a lower limit.

{\it Travel Time for the Ionization Front}:
The radiation front from Cyg OB2 propagates through the ISM between it and the cometary feature at a speed of about 20 km~s$^{-1}$, which is the speed of sound in ionized medium. This gives a simple estimate for the propagation time as $(50 \mathrm{pc})/(20 \mathrm{km~s}^{-1})\approx 2.5$~Myr. This is consistent with a more sophisticated model where we consider the expansion of the Str\"omgren sphere created by the stellar wind in ionized medium \citep{Drn}.

{\it Timescale for Star Formation in the Cometary Feature}: To get an estimate of the timescale for protostar formation in the cometary feature, we note that the spectral index for the six protostars in the cometary feature lies between 0.79 and 2.12, which identifies them as rising-SED protostars or class 0/I \citep{protostars}. According to \citet{Evn}, the lifetimes for protostellar classes 0 and I have values of 0.10 and 0.44 Myr or a maximum lifetime of 0.54 Myr.

{\it Kinematic Timescale for Outflows}: We esimate the kinematic timescale of the outflows as the spatial extent of the outflow divided by the outflow velocity to get $(1\mathrm{~pc})/(0.7\mathrm{~km~s}^{-1})\approx 0.03$ Myr. 

Looking to the future, two other timescales are relevant to describe the evolution of this feature.  

{\it Free-fall Collapse Time}: We calculate the free-fall collapse time for the feature given its mean density: $\sqrt{3/(32\pi G \rho)} \approx 0.3$ Myr.

{\it Photoevaporation Timescale}: We also calculate the rate at which photodissociation will destroy the feature, following the scaling in \citet{sch},
\begin{equation}
    \dot{M}=9.5 \times 10^{-9}\phi_{49}^{1/2}r_{14}^{3/2}d_{17}^{-1}
\end{equation}
where $\phi_{49}=q_0/(10^{49}\mathrm{~photons~s}^{-1})$, $R_{14}$=$R/(10^{14}\mathrm{~cm})$, and $d_{17}$= $d/(10^{17}~\mathrm{cm})$ is the distance between the ionizing source and the feature. Plugging in values from Table \ref{tab:compare}, we estimate the time for the cloud to evaporate due to radiation to be $M/\dot{M} \approx 7 $ Myr. Clearly, the photo-evaporation timescale is very long relative to all other timescales, so the cloud will collapse well before the material is dispersed by radiation.

\begin{figure}
    \centering
    \includegraphics[width=\columnwidth]{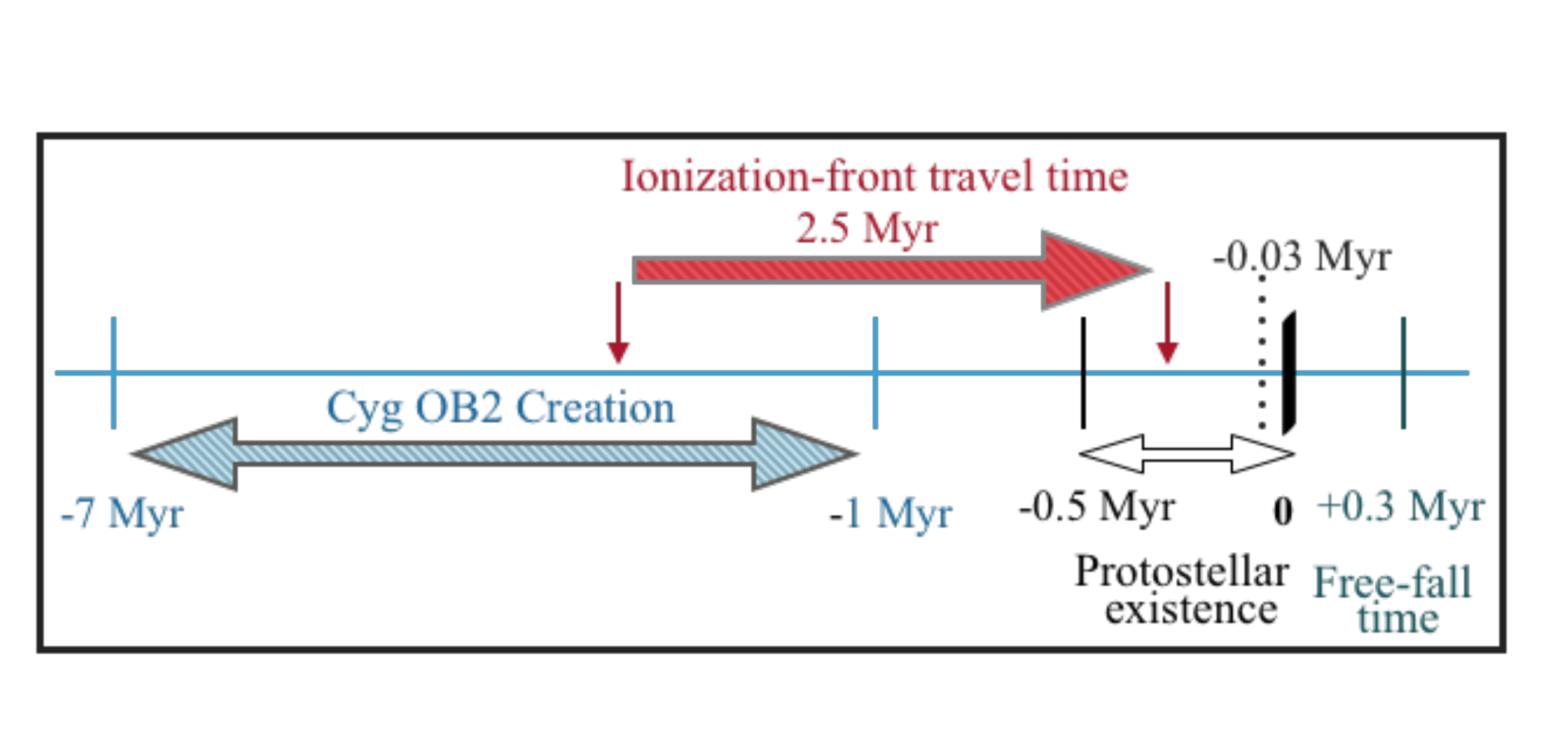}
    \caption{Schematic of the associated timescales for the cometary feature. The dark bold solid vertical line marks the present time. The markers on the left denote time lines in the past. The gray dotted line marks the beginning of the outflows.The solid big red arrow denotes the progress of ionization front in time with fixed length but variable ends marked by small vertical red arrows.  \label{fig:timescale} }
\end{figure}

We lay out these timescales in Figure \ref{fig:timescale}. It is evident that the ionizing radiation front has arrived between 1 Myr ago and the present, which is consistent with the radiation front being originated from Cyg OB2. Ideally, in a triggering scenario, the ionization front would arrive slightly before (to left of) the $-0.5$ Myr mark so its arrival predates the collapse to form the protostars that we see. If the ionization front is not launched until relatively late in the evolution of the Cyg OB2 region (in the past 3 Myr), the front would arrive after stars start to form and there would be no triggering.  However, such timing disagrees with the expected star formation history of the Cyg OB2 region \citep{wright15}.  Thus, the timescales are consistent with a triggering scenario but cannot, by themselves, confirm that the region has been triggered by the ionization front.

\subsection{The Case for Triggering}

This feature shows several indicators that the star formation may have been triggered by the ionizing radation from Cyg OB2. Given $P_\mathrm{ion} > P_\mathrm{mol}$, we are in the domain of radiatively-driven implosion from the pressure in the ionized gas. The illuminated bright rim as observed in H$\alpha$ and radio continuum is evidence of a strong influence of the ionization front at the edge of the cloud. As discussed in sec 4.1, the pillar structure of the cometary feature with the tail pointed towards Cyg OB2 indicates radiation or shock waves coming from OB2, as it is the most likely source present in the vicinity. In addition, we have shown in section 4.3 that the timing is consistent with triggering.\\

This feature is similar to the classic triggered star formation scenario seen in the Eagle nebula.  We compare characteristics of the two regions in Table \ref{tab:compare}. For the Eagle Nebula, total flux emitted by its ionizing source (1.9 pc away) is about 2$\times 10^{50}$ photons~s$^{-1}$ \citep{hester96} which makes the total ionizing flux received at the tip of the pillars $\left(q_0 =Q_0/4\pi d^2\right)$ much higher than that at the tip of the cometary feature, in which case $10^{50.8}$ photons s$^{-1}$ are emitted by the possible ionizing source Cyg OB2 located at a much greater distance (50 pc). However, the gas mass is also much higher for the Eagle Nebula's pillar II. Therefore, this does not rule out the possibility of a collapse in the cometary feature imposed by the radiation front of Cygnus OB2. The relative profile of energy densities in the Eagle Nebula is quite comparable to that in the cometary feature as shown in Table \ref{tab:compare}, which also supports the case for triggering. \\
   
Although, \citet{dale15} clearly demonstrate that even with the above ensemble of circumstantial evidence the case for triggering cannot be made conclusively. Two features argue consistency with triggering as measured in simulations, namely, timing (refer to sec 4.3) and spatial correlation between the locations of the stars and the ionization front. The latter implies that in a large sample of cases of star formation, the distance between the ionizing front and the protostellar objects in a collapsing cloud serves as a (probabilistic) measure of confidence for triggered star formation. This result is shown in Fig. 4 of \citet{dale15}. \\

We estimate that the protostars in the cometary feature have offsets of $<0.2$ pc from the ionized rim. The statistical analysis of simulations given in \citet{dale15} implies that such small separations indicate the protostar formation was likely ($\rm Probability>0.5$) triggered by the ionizing front.  However, without full knowledge of the previous evolution we are unable to make the case conclusively for this individual object.\\

While we cannot conclusively infer the past behaviour, there are some implications for future evolution of the region. The internal support for the region is relatively weak compared to the combined effects of self-gravity and external ionized gas pressure. The free-fall collapse time is short for the global region, particularly when compared to the photo-evaporative timescale. Thus, the global collapse of this region seems inevitable.  However, the region only requires a magnetic field of $70~\mu$G to provide support and mitigate the collapse.  Polarimetric observations of the dust continuum would be able to resolve the geometry and field strength to confirm the possibility of magnetic support.


\section{Summary}
This paper presents new JCMT $^{13}$CO(3-2) and C$^{18}$O(3-2) mapping of a cometary feature in the Cygnus X star forming region.  We use these new data and archival imaging to study the effects of radiative feedback on the cometary feature, evaluating whether the star formation was triggered by the ionization front.   We calculated the properties of the region in detail, including molecular outflows being driven by protostars in the region.  We report the following results:

\begin{itemize}
    \item Using SED fits to the dust emission, we find the cometary feature has a mass of $M=110~\mathrm{M}_\odot$.  Prior observations reveal that the source hosts the formation of at least 6 Class 0/I sources that are seen in the infrared \citep{protostars}.
    \item Assuming LTE, we calculate the column densities of the CO isotopologues using the molecular line data. The fractional abundances of CO isotopologues are consistent with previous studies of molecular clouds with $^{13}\mathrm{CO}/\mathrm{H}_2\approx 10^{-6}$ and $\mathrm{C^{18}O}/\mathrm{H}_2\approx 1.5\times 10^{-7}$.  The CO emission shows some signs of molecular depletion at the centre of the cometary feature.
    \item There are two molecular outflows present in the cometary feature (see Table \ref{tab:outflows}).  The outflowing mass is large ($M\approx 20~\mathrm{M}_\odot$) but the properties of the outflow are comparable to massive outflows found in other systems.  The velocities of the outflow are significantly larger than the escape speed of the clump, so outflowing gas will not provide significant energy or momentum support to the cometary feature.
    \item Using the radiocontinuum and the H$\alpha$ emission, we estimate the conditions in the ionizing gas at the limb of the cometary feature, finding an electron density of $n_e=140\mbox{ cm}^{-3}$ assuming $T_e = 8\,000~\mathrm{ K}$. Given the inventory of massive stars in the region, the most likely source providing the ionizing radiation to illuminate the edge of the cometary feature is the Cygnus OB2 complex of stars.
    \item The combination of external gas pressure and gravitational binding energy dominate the internal thermal pressure and outflow energies, which should prompt the cometary feature to completely collapse on a free-fall timescale (0.3~Myr).  However, the strength and geometry of the magnetic field in the region is unknown and a field with a characteristic strength of $B\gtrsim 70~\mu\mathrm{G}$ could easily support the feature against collapse.  We note that the clump is comparable to the cometary feature seen in M16, where the magnetic field is $B\sim 170~\mu\mathrm{G}$ \citep{pattle18}.
\end{itemize}
After reviewing the physical conditions of the molecular clump and the timescales and sources for a potential triggering scenario, we conclude that a radiatively driven implosion triggering the ongoing star formation is certainly plausible.  However, numerical simulations show that is difficult to evaluate individual, small-scale triggering scenarios and come to a firm conclusion \citep{dale15}.  In the light of those simulations and the detailed study presented here, we regard radiative triggering as the likely cause for the star formation observed here.

\medskip

We acknowledge the support of the Natural Sciences and Engineering Research Council of Canada (NSERC), funding reference numbers RGPIN-2017-03987 and RGPIN 418517. The James Clerk Maxwell Telescope has historically been operated by the Joint Astronomy Centre on behalf of the Science and Technology Facilities Council of the United Kingdom, the National Research Council of Canada and the Netherlands Organisation for Scientific Research. We wish to recognize and acknowledge the significant cultural role and deep reverence that the summit of Maunakea has always had within the indigenous Hawaiian community.  We are truly grateful for the opportunity to conduct observations from this mountain. In addition to software cited in the paper, we also made use of the {\sc astropy} \citep{astropy} and {\sc matplotlib} \citep{matplotlib} software packages. The Dominion Radio Astrophysical Observatory is a National Facility operated by the National Research Council Canada. The Canadian Galactic Plane Survey is a Canadian project with international partners, and is supported by the Natural Sciences and Engineering Research Council of Canada (NSERC).






\bibliographystyle{mnras}
\bibliography{Outflows}{}

\begin{thebibliography}{}
\makeatletter
\relax
\def\mn@urlcharsother{\let\do\@makeother \do\$\do\&\do\#\do\^\do\_\do\%\do\~}
\def\mn@doi{\begingroup\mn@urlcharsother \@ifnextchar [ {\mn@doi@}
  {\mn@doi@[]}}
\def\mn@doi@[#1]#2{\def\@tempa{#1}\ifx\@tempa\@empty \href
  {http://dx.doi.org/#2} {doi:#2}\else \href {http://dx.doi.org/#2} {#1}\fi
  \endgroup}
\def\mn@eprint#1#2{\mn@eprint@#1:#2::\@nil}
\def\mn@eprint@arXiv#1{\href {http://arxiv.org/abs/#1} {{\tt arXiv:#1}}}
\def\mn@eprint@dblp#1{\href {http://dblp.uni-trier.de/rec/bibtex/#1.xml}
  {dblp:#1}}
\def\mn@eprint@#1:#2:#3:#4\@nil{\def\@tempa {#1}\def\@tempb {#2}\def\@tempc
  {#3}\ifx \@tempc \@empty \let \@tempc \@tempb \let \@tempb \@tempa \fi \ifx
  \@tempb \@empty \def\@tempb {arXiv}\fi \@ifundefined
  {mn@eprint@\@tempb}{\@tempb:\@tempc}{\expandafter \expandafter \csname
  mn@eprint@\@tempb\endcsname \expandafter{\@tempc}}}

\bibitem[\protect\citeauthoryear{{Arce} \& {Goodman}}{{Arce} \&
  {Goodman}}{2001}]{arce01}
{Arce} H.~G.,  {Goodman} A.~A.,  2001, \mn@doi [\apj] {10.1086/321334}, \href
  {https://ui.adsabs.harvard.edu/#abs/2001ApJ...554..132A} {554, 132}

\bibitem[\protect\citeauthoryear{{Arce}, {Borkin}, {Goodman}, {Pineda}  \&
  {Halle}}{{Arce} et~al.}{2010}]{arce10}
{Arce} H.~G.,  {Borkin} M.~A.,  {Goodman} A.~A.,  {Pineda} J.~E.,   {Halle}
  M.~W.,  2010, \mn@doi [\apj] {10.1088/0004-637X/715/2/1170}, \href
  {https://ui.adsabs.harvard.edu/#abs/2010ApJ...715.1170A} {715, 1170}

\bibitem[\protect\citeauthoryear{{Astropy Collaboration} et~al.,}{{Astropy
  Collaboration} et~al.}{2013}]{astropy}
{Astropy Collaboration} et~al., 2013, \mn@doi [\aap]
  {10.1051/0004-6361/201322068}, \href
  {https://ui.adsabs.harvard.edu/#abs/2013A&A...558A..33A} {558, A33}

\bibitem[\protect\citeauthoryear{{Barentsen} et~al.,}{{Barentsen}
  et~al.}{2014}]{IPHAS}
{Barentsen} G.,  et~al., 2014, \mn@doi [\mnras] {10.1093/mnras/stu1651}, \href
  {http://adsabs.harvard.edu/abs/2014MNRAS.444.3230B} {444, 3230}

\bibitem[\protect\citeauthoryear{{Comer{\'o}n} \& {Pasquali}}{{Comer{\'o}n} \&
  {Pasquali}}{2012}]{B1}
{Comer{\'o}n} F.,  {Pasquali} A.,  2012, \mn@doi [\aap]
  {10.1051/0004-6361/201219022}, \href
  {http://adsabs.harvard.edu/abs/2012A%26A...543A.101C} {543, A101}

\bibitem[\protect\citeauthoryear{{Condon}, {Cotton}, {Greisen}, {Yin},
  {Perley}, {Taylor}  \& {Broderick}}{{Condon} et~al.}{1998}]{nvss}
{Condon} J.~J.,  {Cotton} W.~D.,  {Greisen} E.~W.,  {Yin} Q.~F.,  {Perley}
  R.~A.,  {Taylor} G.~B.,   {Broderick} J.~J.,  1998, \mn@doi [\aj]
  {10.1086/300337}, \href {http://adsabs.harvard.edu/abs/1998AJ....115.1693C}
  {115, 1693}

\bibitem[\protect\citeauthoryear{{Crutcher}}{{Crutcher}}{2012}]{crutcher}
{Crutcher} R.~M.,  2012, \mn@doi [Annual Review of Astronomy and Astrophysics]
  {10.1146/annurev-astro-081811-125514}, \href
  {https://ui.adsabs.harvard.edu/#abs/2012ARA&A..50...29C} {50, 29}

\bibitem[\protect\citeauthoryear{{Currie}, {Berry}, {Jenness}, {Gibb}, {Bell}
  \& {Draper}}{{Currie} et~al.}{2014}]{STARL}
{Currie} M.~J.,  {Berry} D.~S.,  {Jenness} T.,  {Gibb} A.~G.,  {Bell} G.~S.,
  {Draper} P.~W.,  2014, in Astronomical Data Analysis Software and Systems
  XXIII. p.~391

\bibitem[\protect\citeauthoryear{{Dale}, {Ercolano}  \& {Bonnell}}{{Dale}
  et~al.}{2013}]{Dale}
{Dale} J.~E.,  {Ercolano} B.,   {Bonnell} I.~A.,  2013, \mn@doi [\mnras]
  {10.1093/mnras/sts592}, \href
  {https://ui.adsabs.harvard.edu/#abs/2013MNRAS.430..234D} {430, 234}

\bibitem[\protect\citeauthoryear{{Dale}, {Haworth}  \& {Bressert}}{{Dale}
  et~al.}{2015}]{dale15}
{Dale} J.~E.,  {Haworth} T.~J.,   {Bressert} E.,  2015, \mn@doi [\mnras]
  {10.1093/mnras/stv396}, \href
  {https://ui.adsabs.harvard.edu/#abs/2015MNRAS.450.1199D} {450, 1199}

\bibitem[\protect\citeauthoryear{{Draine}}{{Draine}}{2011}]{Drn}
{Draine} B.~T.,  2011, {Physics of the Interstellar and Intergalactic Medium}

\bibitem[\protect\citeauthoryear{{Evans} Neal~J. et~al.,}{{Evans}
  et~al.}{2009}]{Evn}
{Evans} Neal~J. I.,  et~al., 2009, \mn@doi [The Astrophysical Journal
  Supplement Series] {10.1088/0067-0049/181/2/321}, \href
  {https://ui.adsabs.harvard.edu/#abs/2009ApJS..181..321E} {181, 321}

\bibitem[\protect\citeauthoryear{{Gottschalk}, {Kothes}, {Matthews},
  {Landecker}  \& {Dent}}{{Gottschalk} et~al.}{2012}]{gott12}
{Gottschalk} M.,  {Kothes} R.,  {Matthews} H.~E.,  {Landecker} T.~L.,   {Dent}
  W.~R.~F.,  2012, \mn@doi [\aap] {10.1051/0004-6361/201118600}, \href
  {http://adsabs.harvard.edu/abs/2012A%26A...541A..79G} {541, A79}

\bibitem[\protect\citeauthoryear{{Haffner}, {Reynolds}  \& {Tufte}}{{Haffner}
  et~al.}{1998}]{haffner98}
{Haffner} L.~M.,  {Reynolds} R.~J.,   {Tufte} S.~L.,  1998, \mn@doi [\apjl]
  {10.1086/311449}, \href {http://adsabs.harvard.edu/abs/1998ApJ...501L..83H}
  {501, L83}

\bibitem[\protect\citeauthoryear{{Hester} et~al.,}{{Hester}
  et~al.}{1996}]{hester96}
{Hester} J.~J.,  et~al., 1996, \mn@doi [\aj] {10.1086/117968}, \href
  {https://ui.adsabs.harvard.edu/#abs/1996AJ....111.2349H} {111, 2349}

\bibitem[\protect\citeauthoryear{{Hildebrand}}{{Hildebrand}}{1983}]{Hil}
{Hildebrand} R.~H.,  1983, \qjras, \href
  {http://adsabs.harvard.edu/abs/1983QJRAS..24..267H} {24, 267}

\bibitem[\protect\citeauthoryear{{Hunter}}{{Hunter}}{2007}]{matplotlib}
{Hunter} J.~D.,  2007, \mn@doi [Computing in Science and Engineering]
  {10.1109/MCSE.2007.55}, \href
  {https://ui.adsabs.harvard.edu/#abs/2007CSE.....9...90H} {9, 90}

\bibitem[\protect\citeauthoryear{{Kennicutt} \& {Evans}}{{Kennicutt} \&
  {Evans}}{2012}]{kennicutt12}
{Kennicutt} R.~C.,  {Evans} N.~J.,  2012, \mn@doi [Annual Review of Astronomy
  and Astrophysics] {10.1146/annurev-astro-081811-125610}, \href
  {https://ui.adsabs.harvard.edu/#abs/2012ARA&A..50..531K} {50, 531}

\bibitem[\protect\citeauthoryear{{Klein}, {McKee}  \& {Colella}}{{Klein}
  et~al.}{1994}]{klein94}
{Klein} R.~I.,  {McKee} C.~F.,   {Colella} P.,  1994, \mn@doi [\apj]
  {10.1086/173554}, \href
  {https://ui.adsabs.harvard.edu/#abs/1994ApJ...420..213K} {420, 213}

\bibitem[\protect\citeauthoryear{{Kryukova} et~al.,}{{Kryukova}
  et~al.}{2014}]{protostars}
{Kryukova} E.,  et~al., 2014, \mn@doi [\aj] {10.1088/0004-6256/148/1/11}, \href
  {http://adsabs.harvard.edu/abs/2014AJ....148...11K} {148, 11}

\bibitem[\protect\citeauthoryear{{Lefloch} \& {Lazareff}}{{Lefloch} \&
  {Lazareff}}{1994}]{lefloch94}
{Lefloch} B.,  {Lazareff} B.,  1994, \aap, \href
  {https://ui.adsabs.harvard.edu/#abs/1994A&A...289..559L} {289, 559}

\bibitem[\protect\citeauthoryear{{Makin} \& {Froebrich}}{{Makin} \&
  {Froebrich}}{2018}]{makin18}
{Makin} S.~V.,  {Froebrich} D.,  2018, \mn@doi [The Astrophysical Journal
  Supplement Series] {10.3847/1538-4365/aa8862}, \href
  {https://ui.adsabs.harvard.edu/#abs/2018ApJS..234....8M} {234, 8}

\bibitem[\protect\citeauthoryear{{Mangum} \& {Shirley}}{{Mangum} \&
  {Shirley}}{2015}]{mangum}
{Mangum} J.~G.,  {Shirley} Y.~L.,  2015, \mn@doi [Publications of the
  Astronomical Society of the Pacific] {10.1086/680323}, \href
  {https://ui.adsabs.harvard.edu/#abs/2015PASP..127..266M} {127, 266}

\bibitem[\protect\citeauthoryear{{Molinari} et~al.,}{{Molinari}
  et~al.}{2016}]{higal}
{Molinari} S.,  et~al., 2016, \mn@doi [\aap] {10.1051/0004-6361/201526380},
  \href {https://ui.adsabs.harvard.edu/#abs/2016A&A...591A.149M} {591, A149}

\bibitem[\protect\citeauthoryear{{Pattle} et~al.,}{{Pattle}
  et~al.}{2018}]{pattle18}
{Pattle} K.,  et~al., 2018, \mn@doi [\apj] {10.3847/2041-8213/aac771}, \href
  {https://ui.adsabs.harvard.edu/#abs/2018ApJ...860L...6P} {860, L6}

\bibitem[\protect\citeauthoryear{{Rygl} et~al.,}{{Rygl} et~al.}{2012}]{rygl12}
{Rygl} K.~L.~J.,  et~al., 2012, \mn@doi [\aap] {10.1051/0004-6361/201118211},
  \href {http://adsabs.harvard.edu/abs/2012A%26A...539A..79R} {539, A79}

\bibitem[\protect\citeauthoryear{{Schneider} et~al.,}{{Schneider}
  et~al.}{2016}]{sch}
{Schneider} N.,  et~al., 2016, \mn@doi [\aap] {10.1051/0004-6361/201628328},
  \href {https://ui.adsabs.harvard.edu/#abs/2016A&A...591A..40S} {591, A40}

\bibitem[\protect\citeauthoryear{{Sternberg}, {Hoffmann}  \&
  {Pauldrach}}{{Sternberg} et~al.}{2003}]{sternberg03}
{Sternberg} A.,  {Hoffmann} T.~L.,   {Pauldrach} A.~W.~A.,  2003, \mn@doi
  [\apj] {10.1086/379506}, \href
  {http://adsabs.harvard.edu/abs/2003ApJ...599.1333S} {599, 1333}

\bibitem[\protect\citeauthoryear{{Taylor} et~al.,}{{Taylor}
  et~al.}{2003}]{cgps}
{Taylor} A.~R.,  et~al., 2003, \mn@doi [\aj] {10.1086/375301}, \href
  {http://adsabs.harvard.edu/abs/2003AJ....125.3145T} {125, 3145}

\bibitem[\protect\citeauthoryear{{Wendker}, {Higgs}  \& {Landecker}}{{Wendker}
  et~al.}{1991}]{wendker91}
{Wendker} H.~J.,  {Higgs} L.~A.,   {Landecker} T.~L.,  1991, \aap, \href
  {http://adsabs.harvard.edu/abs/1991A%26A...241..551W} {241, 551}

\bibitem[\protect\citeauthoryear{{Wilson} \& {Rood}}{{Wilson} \&
  {Rood}}{1994}]{wilson94}
{Wilson} T.~L.,  {Rood} R.,  1994, \mn@doi [Annual Review of Astronomy and
  Astrophysics] {10.1146/annurev.aa.32.090194.001203}, \href
  {https://ui.adsabs.harvard.edu/#abs/1994ARA&A..32..191W} {32, 191}

\bibitem[\protect\citeauthoryear{{Wright}, {Drew}  \& {Mohr-Smith}}{{Wright}
  et~al.}{2015}]{wright15}
{Wright} N.~J.,  {Drew} J.~E.,   {Mohr-Smith} M.,  2015, \mn@doi [\mnras]
  {10.1093/mnras/stv323}, \href
  {http://adsabs.harvard.edu/abs/2015MNRAS.449..741W} {449, 741}

\makeatother
\end{thebibliography}
    	






\bsp	
\label{lastpage}
\end{document}